%% file: main.tex
\documentclass{article}
\usepackage[margin=1in]{geometry}
\usepackage[utf8]{inputenc}
\usepackage{graphicx}
\usepackage[citestyle=ieee]{biblatex}
\usepackage{amsmath}
\usepackage{amssymb}
\usepackage{amsthm}
\usepackage{tikz}
\usepackage{pgfplots}
\usepackage{pgfplotstable}
\usepackage[short]{optidef}
\usepackage{subcaption}
\usepackage{pgf}
\usepackage{algorithm}
\usepackage{algorithmic}

\newtheorem{theorem}{Theorem}
\newtheorem{assumption}{Assumption}
\newtheorem{lemma}{Lemma}
\newtheorem{definition}{Definition}

\usetikzlibrary{arrows, angles, quotes, external, fit, positioning, calc, backgrounds}
\pgfplotsset{compat=1.15}
\usepgfplotslibrary{groupplots}
\DeclareUnicodeCharacter{2212}{-}

\newcommand\blfootnote[1]{%
  \begingroup
  \renewcommand\thefootnote{}\footnote{#1}%
  \addtocounter{footnote}{-1}%
  \endgroup
}

\title{Optimal model-based trajectory planning with static polygonal constraints}
\author{Andreas B. \ Martinsen, Anastasios M. \ Lekkas, and S{\'e}bastien\ Gros}
\date{}

\begin{document}

\maketitle
\blfootnote{The authors are with the Department of Engineering Cybernetics, Norwegian University of Science and Technology (NTNU), O. S. Bragstads plass 2D, NO-7491 Trondheim, Norway E-mails:\{andreas.b.martinsen, anastasios.lekkas, sebastien.gros\}@ntnu.no}

\begin{abstract}
    \input{content/abstract.tex}
\end{abstract}

\input{content/introduction.tex}

\input{content/method.tex}

\input{content/example.tex}

\input{content/conclusion.tex}


\printbibliography 

\end{document}

%% file: content/abstract.tex
The main contribution of this paper is a novel method for planning globally optimal trajectories for dynamical systems subject to polygonal constraints. The proposed method is a hybrid trajectory planning approach, which combines graph search, i.e. a discrete roadmap method, with convex optimization, i.e. a complete path method. Contrary to past approaches, which have focused on using simple obstacle approximations, or sub-optimal spatial discretizations, our approach is able to use the exact geometry of polygonal constraints in order to plan optimal trajectories. The performance and flexibility of the proposed method is evaluated via simulations by planning distance-optimal trajectories for a Dubins car model, as well as time-, distance- and energy-optimal trajectories for a marine vehicle.

%% file: content/introduction.tex
\section{Introduction} \label{sec:intoduction}
In robotics, motion planning is the process of finding a sequence of valid configurations, which move the robot safely from some initial configuration to a goal configuration. To be successful in the real world, the motion planner must be able to consider a variety of constraints such as \textit{environment constraints}, including static and dynamic obstacles, and  \textit{differential constraints}, which arise from the system kinematics and dynamics and are modeled with differential equations. Due to a potentially large number of obstacles, actuators, as well as complex kinematics and dynamics, motion planning is in general a difficult problem that has led to a wide range of methods and a vast literature. 

Trajectory planning pertains to finding a time-parametric continuous sequence of configurations, called a trajectory, which is obstacle-free and satisfies the differential constraints (i.e. a feasible trajectory). Optimal trajectory planning has the additional task of finding the ''best'' feasible trajectory with respect to some performance measure, such as minimum energy, distance or time. The requirement of optimality is in general very demanding computationally since it requires an exhaustive search over the state space. One of many ways to categorize motion planning methods is to distinguish between \textit{roadmap methods} and \textit{complete path methods} \cite{wolek2017model, bitar2018energy, lavalle2006planning}.

The main goal of roadmap methods is to find a sequence of waypoints, which, when connected, result in an obstacle-free piecewise-linear path. The path can then be smoothed and turned into a feasible trajectory that complies with the vehicle dynamics. Roadmap methods can be further split into two distinct categories, namely, combinatorial methods and sampling-based methods. Combinatorial methods, divide the continuous space into structures that capture all spatial information needed to solve the motion planning using simple graph search algorithms. For many complex problems however, combinatorial methods may not be computationally feasible. For these problems, sampling based methods are often used instead. Sampling based methods, rely on using randomly sampled subset of states or actions. This creates a randomly sampled discretization of the continuous search space, and hence limits the computational complexity at the cost of accuracy and completeness of the discretization. Some notable combinatorial methods include coarse planning with path smoothing, in where a mesh, grid or potential field is used to plan a course path \cite{kallmann2005path, hart1968formal, candeloro2017voronoi, barraquand1992numerical}, and then a method using curve segments, splines or motion primitives is used to refine the trajectory \cite{lekkas2013continuous, jacobs1993planning, fleury1995primitives, judd2001spline, pan2012collision, lekkas2014integral, bottasso2008path}. Notable sampling based methods include probabilistic roadmap (PRM) \cite{kavraki1996probabilistic}, rapidly-exploring random tree (RRT) \cite{lavalle1998rapidly, lavalle2001randomized, karaman2010incremental}, and Random-walk planners \cite{carpin2005motion, carpin2005merging}

Complete path methods on the other hand, produce a continuous parameterized trajectory by explicitly taking into account the motion equations of the robot and the full continuous search space. As a result, these methods generate a trajectory that is both obstable-free and feasible, without further need of refinement/smoothing. Most complete path methods rely on some form of mathematical optimization. For some simple problems an analytical solution exists, as is the case for Dubins paths \cite{dubins1957curves} and Reeds-Shepp \cite{reeds1990optimal}. In general, however, researchers must resort to numerical optimization, where handling complex constraints is challenging  and getting stuck in local optima is not uncommon. Notable numerical methods include dynamic programming \cite{bellman1966dynamic}, particle swarm optimization (PSO) \cite{eberhart1995particle, kennedy1997discrete}, shooting methods \cite{bock1984multiple}, which are based on simulation, collocation methods \cite{hargraves1987direct}, which are based on function approximation of low-level polynomials, and pseudospectral methods \cite{fahroo2002direct}, which are based on function approximation of high-level polynomials.



In this paper we consider the problem of optimal motion planning for a particle-like vehicle, moving on a 2D surface with polygonal obstacles. To this end, we introduce a hybrid method, which combines graph search on a pre-computed mesh, with convex optimization for path refinement. The proposed method allows for planning a globally optimal trajectory for a dynamical system subject to static polygonal constraints. The main contributions is this paper is how we combine hybrid planning with polygonal constraints and triangulation based spatial discretization.
With hybrid planning, we combine both roadmap and complete path methods. Contrary to other hybrid methods such as \cite{ljungqvist2017lattice, bergman2020optimization, bitar2019warm}, where initial trajectories are planned using motion primitives and state space discretizations, and refined using numerical optimization, our method employs an iterative approach of planning and refinement.
Polygonal constraints allow for complex constraints to be used in the planning algorithm. Very few optimization-based planning methods exist that are able to handle these types of constraints. Existing methods often lead to computationally expensive mixed integer optimization problems \cite{blackmore2011chance}, rely on using inner approximations of the free space \cite{schoels2020ciao, martinsen2019autonomous}, or non-convex elliptical approximations \cite{bitar2018energy}. 
Our method relies on using a triangulation of the environment, similar to \cite{kallmann2005path, yan2008path} but instead of straight-line paths, it plans the path as a polynomial spline, similar to \cite{mercy2017spline}. Combining the above concepts, our proposed method is able to efficiently plan globally optimal trajectories for a dynamical system subject to static polygonal constraints. 

The rest of the paper is organized as follows: Section \ref{sec:method} outlines the method. Section \ref{sec:example} shows examples of distance-optimal paths for a simple kinematic car, as well as time-, distance- and energy-optimal paths for an unmanned surface vehicle. Finally Section \ref{sec:conclusion} concludes the paper.

%% file: content/method.tex
\section{Method} \label{sec:method}
\begin{figure}
    \centering
    \begin{subfigure}{0.25\linewidth}
        \resizebox{\linewidth}{!}{\input{article/figures/algorithm_polygons.tikz}}
        \caption{Obstacles}
    \end{subfigure}
    ~
    \begin{subfigure}{0.25\linewidth}
        \centering
        \resizebox{\linewidth}{!}{\input{article/figures/algorithm_triangulation.tikz}}
        \caption{Triangulation}
    \end{subfigure}
    ~
    \begin{subfigure}{0.25\linewidth}
        \centering
        \resizebox{\linewidth}{!}{\input{article/figures/algorithm_adjacency.tikz}}
        \caption{Adjacency graph}
    \end{subfigure}
    \\
    \begin{subfigure}{0.25\linewidth}
        \centering
        \resizebox{\linewidth}{!}{\input{article/figures/algorithm_spline.tikz}}
        \caption{Triangle selection}
    \end{subfigure}
    ~
    \begin{subfigure}{0.25\linewidth}
        \centering
        \resizebox{\linewidth}{!}{\input{article/figures/algorithm_refinement.tikz}}
        \caption{Optimized trajectory}
    \end{subfigure}
        ~
    \begin{subfigure}{0.25\linewidth}
        \centering
        \resizebox{\linewidth}{!}{\input{article/figures/algorithm_final.tikz}}
        \caption{Final trajectory}
    \end{subfigure}
    \caption{Given polygonal obstacles (a), the proposed algorithm finds the trajectory by creating a triangulation (b) and adjacency graph (c). Iteratively exploring different triangle sequences (d) where the refined trajectory is optimized as a spline (e). The exploration is performed until the goal is reached (f).}
    \label{fig:algorithm_sequence}
\end{figure}
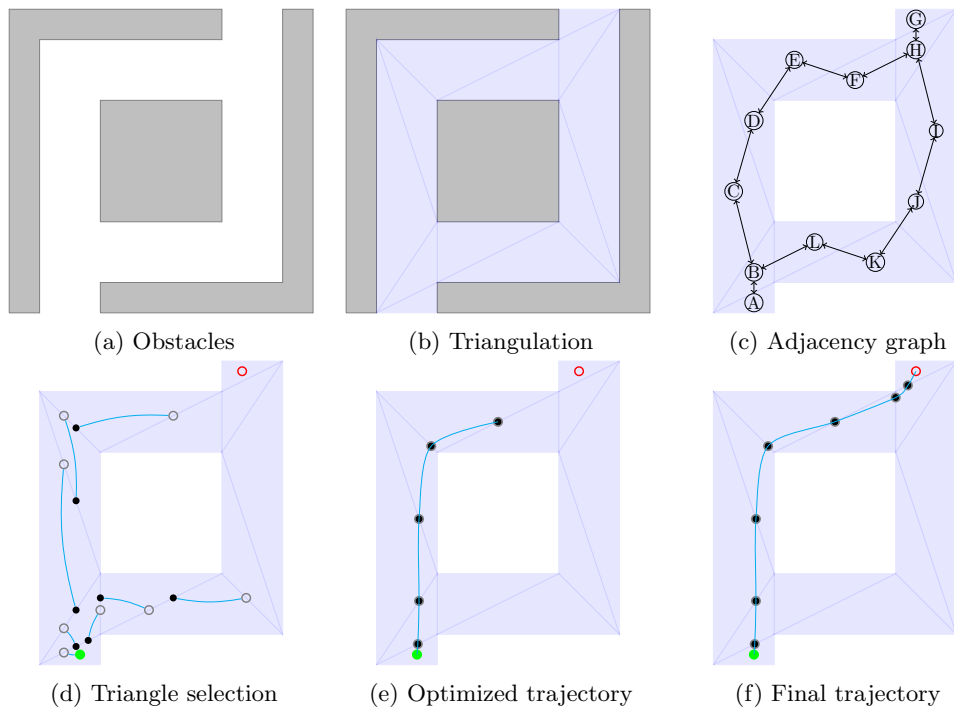

The problem that we aim to solve in this paper, is that of planning optimal trajectories for dynamical systems in environments with static polygonal constraints. The proposed method is able to compute optimal time parameterized state trajectories:
\begin{equation*}
    \boldsymbol{x}(t), \quad t \in [t_0, t_f],
\end{equation*}
which connect some initial state $\boldsymbol{x}_0$ and final goal state $\boldsymbol{x}_f$ such that:
\begin{equation*}
    \boldsymbol{x}(t_0) = \boldsymbol{x}_0, \quad \boldsymbol{x}(t_f) = \boldsymbol{x}_f.
\end{equation*}
The trajectory is generated such that it satisfies the continuous time dynamics and kinematics of a given dynamical system on the form:
\begin{equation*}
    \dot{\boldsymbol{x}} = f(\boldsymbol{x}, \boldsymbol{u}),
\end{equation*}
which in general may be nonlinear and have additional constraints on the states and actions. The optimized trajectory, is found such that it avoids polygonal spatial constraints that are present in the environment. This is ensured by having the path travel through a sequence of neighbouring triangles $\mathcal{T}_i$, with the sequence denoted $[\mathcal{T}_0, \mathcal{T}_1, \dots \mathcal{T}_N]$, where the interior of each triangle is collision free. The proposed method for solving this problem can be divided into three distinct stages. 
\begin{enumerate}
    \item \textbf{Triangulation and adjacency graph} is the first stage, where a triangulation of the environment is generated based on the polygonal constraints (Figure \ref{fig:algorithm_sequence}b), and an adjacency graph is calculated based on neighbouring triangles (Figure \ref{fig:algorithm_sequence}c). 
    
    \item \textbf{Graph search} is the second phase, where a graph search algorithm is used to explore possible sequences of triangles in the triangulation (Figure \ref{fig:algorithm_sequence}c). 
    
    \item \textbf{Trajectory refinement} is the third phase, where a continuous trajectory is generated and optimized within the confinement of a sequence of triangles (Figure \ref{fig:algorithm_sequence}d and \ref{fig:algorithm_sequence}e).
\end{enumerate}

\subsection{Triangulation and adjacency graph}
In this step, the objective is to generate a triangulation of the environment, given polygonal spatial constraints. The resulting triangulation must include the edges of the polygons, which is referred to as \textit{constrained triangulation}. The reason for segmenting the environment into triangles in this way, is that any triangle in this type of triangulation, is either fully inside of the polygonal constraint, or fully outside of the polygonal constraint. This results in an exact, and efficient decomposition of the environment. We can then use the triangles that are fully outside of the polygonal constraints in order to plan a sequence of triangles for the trajectory to pass through, which is guaranteed to be collision free. 

In this work, the triangulation that we use, is a Constrained Delaunay Triangulation (CDT) \cite{chew1989constrained}. A regular Delaunay triangulation (DT) \cite{preparata1985computational} will maximize the minimum angle of all the angles of the triangles in the triangulation, and hence tend to avoid sliver triangles. With CDT, certain segments are forced into the triangulation. This is necessary in order to ensure that the triangles of the triangulation are either fully inside the polygonal spatial constraints, or fully outside the spatial constraints. For the spatial constraints in Figure \ref{fig:algorithm_sequence}a, a constraint triangulation is given in Figure \ref{fig:algorithm_sequence}b.

After the triangulation is created, an adjacency graph is computed by connecting neighbouring triangles of the triangulation, where two triangles are considered neighbours if they share an edge. An illustration is shown in Figure \ref{fig:algorithm_sequence}c. The triangulation and adjacency graph are then used in the next phase for exploring and planning sequences of neighbouring triangles.


\subsection{Graph search}
Graph search can in general only be used for planning in discrete environments. In order to extend it to the continuous domain, we propose using a trajectory refinement strategy, where the graph search is performed by planning a sequence of neighbouring triangles $[\mathcal{T}_0, \mathcal{T}_1, \dots \mathcal{T}_N]$, and a continuous time parameterized trajectory $\boldsymbol{x}(t)$, is planned within the constraints of the sequence of triangles.

Given a CTD, we can construct a graph, where each node represents a triangle, and edges are given by neighbouring triangles, this is illustrated in Figure \ref{fig:algorithm_sequence}c. The goal of the graph search is to plan a sequence of triangles $[\mathcal{T}_0, \dots \mathcal{T}_N]$, which optimizes a desired performance measure. In our case the goal is to optimize a time parameterized path integral on the form:
\begin{equation}
    \int_{t_{0}}^{t_{f}} J(\cdot) d\tau,
\end{equation}
where $J(\cdot)$ is a non-negative instantaneous cost. Given an initial starting point $\boldsymbol{x}_0$, the proposed graph search method, works by staring with the initial triangle sequence $[\mathcal{T}_0]$, such that $\boldsymbol{x}_0 \in \mathcal{T}_0$. It then iteratively extending the sequence of triangles $[\mathcal{T}_0, \dots \mathcal{T}_{N - 1}]$, by adding new neighbouring triangels $\mathcal{T}_{N}$. This is performed until a feasible sequence of triangles $[\mathcal{T}_0, \dots \mathcal{T}_{N}]$, connecting the initial state $\boldsymbol{x}_0$ and final goal state $\boldsymbol{x}_f$, is found, and a termination condition is met. The order in which potential sequences are extended, is determined by a heuristics based lower bound on the path integral. This ensures that the potentially best paths are explored first, and hence reducing the number of triangle sequences that need to be explored.

\subsection{Trajectory refinement}
\begin{figure}
    \centering
    \input{article/figures/triangle_path.tikz}
    \caption{The trajectory ($A \rightarrow C$) through two triangles can be planned as the trajectory through each individual triangle (\textcolor{red}{$A \rightarrow B$} and \textcolor{blue}{$B \rightarrow C$}), constrained to meeting somewhere along the neighbouring edge.}
    \label{fig:trinagle_path}
\end{figure}
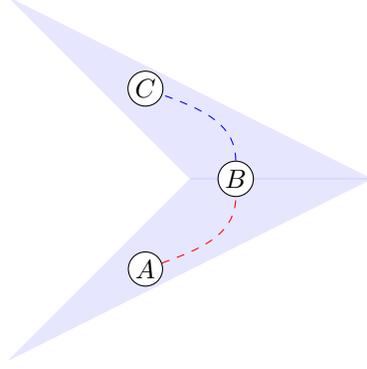

In order to plan a continuous trajectory in an area divided into triangles, we can observe that the trajectory is constrained by the edge through which it enters, and the edge through which it leaves any given triangle. The point at which it leaves and enters a triangle is also the point at which the trajectory enters and leaves its neighbours respectively. It is therefore possible to plan a refined trajectory through each triangle, with a given entrance and exit point along the triangle boundary (see Figure \ref{fig:trinagle_path}). This means that the final optimal trajectory, which may travel through a non-convex polygon, consists of trajectory segments constrained to lie within individual convex triangles.

Given a dynamical system on the form:
\begin{equation}
    \dot{\boldsymbol{x}} = f(\boldsymbol{x}, \boldsymbol{u}),
\end{equation}
where $\boldsymbol{x}$ is the state vector, and $\boldsymbol{u}$ is the control vector. The optimal trajectory through a sequence of neighbouring triangles, denoted $[\mathcal{T}_0, \mathcal{T}_1, \dots \mathcal{T}_N]$, can be written as the following optimization problem.
\begin{subequations} \label{eq:optimization:path_refinement}
\begin{align}
    V(\boldsymbol{x}_0, [\mathcal{T}_0, \mathcal{T}_1, \dots \mathcal{T}_N]) = \min_{\boldsymbol{x}, \boldsymbol{u}, t} \quad & \sum_{i = 0}^{N} \int_{t_{i}}^{t_{i + 1}} J(\boldsymbol{x}, \boldsymbol{u}, \tau) d\tau \\
    \text{s.t.} \quad
        & \dot{\boldsymbol{x}} = f(\boldsymbol{x}, \boldsymbol{u}), \\
        & \boldsymbol{x}(t) \in \mathcal{T}_i \quad \forall t \in [t_{i}, t_{i + 1}]\\ 
        & \boldsymbol{x}(t_0) = \boldsymbol{x}_0. 
\end{align}
\end{subequations}
In the above optimization problem, (\ref{eq:optimization:path_refinement}b) ensures the trajectory is feasible with respect to the model, (\ref{eq:optimization:path_refinement}c) ensures each trajectory segment lies within its respective triangle, and (\ref{eq:optimization:path_refinement}d) gives the initial conditions for the optimization problem. Using the above formulation, we note that in the graph-search phase, the optimization problem is built by iteratively adding triangles to the triangle sequence $[\mathcal{T}_0, \mathcal{T}_1, \dots \mathcal{T}_N]$, and hence extending the horizon $N$.

\subsection{Complete method}

Given a trajectory $\boldsymbol{x}(t)$, starting at $\boldsymbol{x}_0$, and ending at $\boldsymbol{x}_f$, and going through a sequence of triangles $[\mathcal{T}_0, \mathcal{T}_1, \dots \mathcal{T}_N]$, we can define the value function of the sequence as the value that minimizes the cost along the optimal trajectory through the sequence of triangles, with fixed start and endpoint:
\begin{subequations} \label{eq:optimization:Q}
\begin{align}
    Q(\boldsymbol{x}_0, [\mathcal{T}_0, \mathcal{T}_1, \dots \mathcal{T}_N], \boldsymbol{x}_f) = \min_{\boldsymbol{x}, \boldsymbol{u}, t} \quad & \sum_{i = 0}^{N} \int_{t_{i}}^{t_{i + 1}} J(\boldsymbol{x}, \boldsymbol{u}, \tau) d\tau \\
    \text{s.t.} \quad
        & \dot{\boldsymbol{x}} = f(\boldsymbol{x}, \boldsymbol{u}), \\
        & \boldsymbol{x}(t) \in \mathcal{T}_i \quad \forall t \in [t_i, t_{i + 1}]\\ 
        & \boldsymbol{x}(t_0) = \boldsymbol{x}_0 \\
        & \boldsymbol{x}(t_{N+1}) = \boldsymbol{x}_f. \label{eq:optimization:Q:constraint:terminal}
\end{align}
\end{subequations}
Note, that this is the same optimization problem as in \eqref{eq:optimization:path_refinement}, but with the addition of the terminal constraint in \eqref{eq:optimization:Q:constraint:terminal}. Using this, we can get the result in Lemma \ref{le:value_function:fixed_vs_free}.

\begin{lemma}\label{le:value_function:fixed_vs_free}
The fixed endpoint value function $Q(\cdot)$ will always be lower bounded by the free endpoint value function $V(\cdot)$:
\begin{equation}
     Q(\boldsymbol{x}_0, [\mathcal{T}_0, \dots \mathcal{T}_N], \boldsymbol{x}_f) \geq  V(\boldsymbol{x}_0, [\mathcal{T}_0, \dots \mathcal{T}_N])
\end{equation}

\begin{proof}
The free endpoint value function $V(\cdot)$ where $\boldsymbol{x}_f$ is free can be expressed in terms of minimizing the fixed endpoint value function $Q(\cdot)$ as follows:
\begin{equation}
\begin{aligned}
    V(\boldsymbol{x}_0, [\mathcal{T}_0, \dots \mathcal{T}_N]) 
    &= \min_{\boldsymbol{x}_f \in  \mathcal{T}_N} Q(\boldsymbol{x}_0, [\mathcal{T}_0, \dots \mathcal{T}_N], \boldsymbol{x}_f) \\
    &\leq  Q(\boldsymbol{x}_0, [\mathcal{T}_0, \dots \mathcal{T}_N], \boldsymbol{x}_f) \: \forall  \boldsymbol{x}_f \in \mathcal{T}_N
\end{aligned}
\end{equation}
\end{proof}
\end{lemma}

In order to determine the optimality of a sequence of triangles, we need to show that extending the sequence will not lower the cost of the trajectory. Using the value function definitions in \eqref{eq:optimization:path_refinement} and \eqref{eq:optimization:Q}, and the following assumption, we get the result in Lemma \ref{le:monotonic_value_function}.

\begin{assumption}\label{as:non-negative_cost}
The cost function $J(\cdot) \geq 0$ is a non-negative function. Meaning the integral of the cost can not decrease along the path.
\end{assumption}

\begin{lemma}\label{le:monotonic_value_function}
Given Assumption \ref{as:non-negative_cost}, the value function $ V(\boldsymbol{x}_0, [\mathcal{T}_0, \dots \mathcal{T}_N])$ is monotonically increasing with respect to the length of the sequence of triangles. 

\begin{proof}
\begin{equation*}
\begin{aligned}
 V(\boldsymbol{x}_0, [\mathcal{T}_0, \dots \mathcal{T}_{N}])
 &= Q(\boldsymbol{x}_0, [\mathcal{T}_0, \dots \mathcal{T}_{N - 1}], \boldsymbol{x}_N) + V(\boldsymbol{x}_N, [\mathcal{T}_N])\\
 &\geq V(\boldsymbol{x}_0, [\mathcal{T}_0, \dots \mathcal{T}_{N - 1}]) + V(\boldsymbol{x}_N, [\mathcal{T}_N]) \\
 &\geq V(\boldsymbol{x}_0, [\mathcal{T}_0, \dots \mathcal{T}_{N - 1}]) 
\end{aligned}
\end{equation*}
\end{proof}
\end{lemma}

\begin{definition}[Triangle sequence completeness]
    We say that a sequence of triangles $[\mathcal{T}_0, \dots \mathcal{T}_N]$ is \underline{complete} if the initial state is within the initial triangle $\boldsymbol{x}_0 \in \mathcal{T}_0$, and the final goal state is in the final triangle $\boldsymbol{x}_f \in \mathcal{T}_N$. Similarly, a sequence is \underline{incomplete} if the initial state is within initial triangle $\boldsymbol{x}_0 \in \mathcal{T}_0$, and the final goal state is not within the last triangle  $\boldsymbol{x}_f \notin \mathcal{T}_N$.
\end{definition}

When searching sequences of triangles, it is useful to be able to approximate bounds on the cost to go, if the sequence is incomplete. In order to do this, we are using an admissible heuristic function $h(\boldsymbol{x}, \boldsymbol{x}_f)$ to estimate the cost to go from some state $\boldsymbol{x}$ to the terminal goal state $\boldsymbol{x}_f$. Using the heuristic, we can define the following function:
\begin{equation}
\begin{aligned}
    \underline{Q}(\boldsymbol{x}_0, [\mathcal{T}_0,\dots \mathcal{T}_{M-1}], \boldsymbol{x}_f) 
    &= V(\boldsymbol{x}_0, [\mathcal{T}_0,\dots \mathcal{T}_{M-1}]) \\
    &+ h(\boldsymbol{x}_M, \boldsymbol{x}_f), \: \boldsymbol{x}_M = \boldsymbol{x}(t_M)
\end{aligned}
\end{equation}
which is a lower bound on possible complete sequences of triangles, that extend from an incomplete sequence. This result is summed up in Lemma \ref{le:cost_lower_bound}.

\begin{assumption}\label{as:admissible_heuristic} The heuristic function $h(\boldsymbol{x}_M, \boldsymbol{x}_f)$ is admissible. Hence the the heuristic will always underestimate the true cost or value function for any feasible sequence of triangles $[\mathcal{T}_M , \dots \mathcal{T}_N]$. 
\begin{equation*}
    h(\boldsymbol{x}_M, \boldsymbol{x}_f ) \leq Q(\boldsymbol{x}_M, [\mathcal{T}_M , \dots \mathcal{T}_N], \boldsymbol{x}_f),
\end{equation*}
\end{assumption}

\begin{lemma}\label{le:cost_lower_bound}
Given Assumption \ref{as:admissible_heuristic} and a triangle sequence $[\mathcal{T}_0, \dots \mathcal{T}_M , \dots \mathcal{T}_N]$, we have the following lower bound on the trajectory cost:
\begin{equation}
\begin{aligned}
     Q(\boldsymbol{x}_0, [\mathcal{T}_0,\dots \mathcal{T}_N], \boldsymbol{x}_f) \geq
     \underbrace{V(\boldsymbol{x}_0, [\mathcal{T}_0,\dots \mathcal{T}_{M-1}]) + h(\boldsymbol{x}_M, \boldsymbol{x}_f)}_{:= \: \underline{Q}(\boldsymbol{x}_0, [\mathcal{T}_0,\dots \mathcal{T}_{M-1}], \boldsymbol{x}_f)}
\end{aligned}
\end{equation}
where $\boldsymbol{x}_M = \boldsymbol{x}(t_M)$ is the end of the optimal free endpoint trajectory given by $V(\boldsymbol{x}_0, [\mathcal{T}_0,\dots \mathcal{T}_{M-1}]$.

\begin{proof}
\begin{equation*}
\begin{aligned}
    Q(\boldsymbol{x}_0, [\mathcal{T}_0, \dots \mathcal{T}_N], \boldsymbol{x}_f)
    &= Q(\boldsymbol{x}_0, [\mathcal{T}_0, \dots \mathcal{T}_{M-1}], \boldsymbol{x}_M) \\
    &+ Q(\boldsymbol{x}_M, [\mathcal{T}_M , \dots \mathcal{T}_N], \boldsymbol{x}_f) \\
    &\geq V(\boldsymbol{x}_0, [\mathcal{T}_0, \dots \mathcal{T}_{M-1}]) \\
    &+ Q(\boldsymbol{x}_M, [\mathcal{T}_M , \dots \mathcal{T}_N], \boldsymbol{x}_f) \\
    &\geq V(\boldsymbol{x}_0, [\mathcal{T}_0, \dots \mathcal{T}_{M-1}]) \\
    &+ h(\boldsymbol{x}_M, \boldsymbol{x}_f) \\
    &= \underline{Q}(\boldsymbol{x}_0, [\mathcal{T}_0,\dots \mathcal{T}_{M-1}], \boldsymbol{x}_f)
\end{aligned}
\end{equation*}
\end{proof}
\end{lemma}

Given the result from Lemma \ref{le:cost_lower_bound}, where we have a lower bound $\underline{Q}(\cdot)$ for completing an incomplete sequence of triangles, we can use this to determine if completing an incomplete path will result in a complete sequence with a lower value $Q(\cdot)$, then some other completes sequence. This is summed up in Theorem \ref{th:optimal_termination}.

\begin{theorem} \label{th:optimal_termination}
Given a complete sequence of triangles $\mathcal{S}^* = [\mathcal{T}_0, \dots \mathcal{T}_N]$, and an incomplete sequence $\mathcal{S}'$ satisfying:
\begin{equation} \label{eq:termination_condition}
    Q(\boldsymbol{x}_0, \mathcal{S}^*, \boldsymbol{x}_f) \leq \underline{Q}(\boldsymbol{x}_0, \mathcal{S}', \boldsymbol{x}_f),
\end{equation}
Then completing the incomplete sequence $\mathcal{S}'$ can not result in a trajectory with a lower value $Q(\cdot)$ then the sequence $\mathcal{S}^*$.

\begin{proof}
From Lemma \ref{le:cost_lower_bound}, we have that extending any incomplete sequence $\mathcal{S}'$ to a complete sequence $\mathcal{S}$ will result in a higher cost, i.e:
\begin{equation*}
    \underline{Q}(\boldsymbol{x}_0, \mathcal{S}', \boldsymbol{x}_f) 
    \leq Q(\boldsymbol{x}_0, \mathcal{S}, \boldsymbol{x}_f).
\end{equation*}
Given the condition in \eqref{eq:termination_condition}, we get the following result:
\begin{equation*}
\begin{aligned}
    Q(\boldsymbol{x}_0, \mathcal{S}^*, \boldsymbol{x}_f)
    \leq \underline{Q}(\boldsymbol{x}_0, \mathcal{S}', \boldsymbol{x}_f) 
    \quad \Rightarrow \quad \\
    Q(\boldsymbol{x}_0, \mathcal{S}^*, \boldsymbol{x}_f)
    \leq Q(\boldsymbol{x}_0, \mathcal{S}, \boldsymbol{x}_f).
\end{aligned}
\end{equation*}
This means that all sequences $\mathcal{S}$ that can result from the incomplete sequences $\mathcal{S}'$ will have higher cost then the optimal sequence $\mathcal{S}^*$ if  \eqref{eq:termination_condition} holds.
\end{proof}
\end{theorem}

Using the refined trajectory cost $V(\cdot)$, heuristic admissible cost $h(\cdot)$ and the search termination conditions given Theorem \ref{th:optimal_termination}, we can derive the complete trajectory planning Algorithm \ref{alg:planning}. Where at each iteration, the trajectory is expanded into the triangle that minimizes the cost lower bound $\underline{Q}(\cdot)$. Until a complete sequence of triangles $\mathcal{S}^*$ is found, for which the termination condition in \eqref{eq:termination_condition} is true for all sequences $\mathcal{S}'$, in the list of sequences to be searched ($open\_list$). From Theorem \ref{th:optimal_algorithm} we can show that the proposed algorithm will find the optimal sequence of triangles, and hence the globally optimal trajectory, under the assumption that the resulting optimization problem is convex. This is the case if the dynamical system results in convex constraints, as the spatial constraints will be convex due to the triangulation. 

\begin{lemma} \label{le:open_list_sub-sequences}
In Algorithm \ref{alg:planning}, the list of sequences to be searched ($open\_list$) will always contain a sub-sequence $\mathcal{S}'$ of any possible complete path $\mathcal{S}$

\begin{proof}
Algorithm \ref{alg:planning}, changes the $open\_list$ by iterative removing incomplete sequences, and adding all feasible sequences that can be extended by one triangle from the sequence that is removed. Since any possible complete path must be extended from the sequence only containing the initial triangle $\mathcal{T}_0$. Then the list of sequences to be searched ($open\_list$) will always contain a sub-sequence of any possible complete path.
\end{proof}
\end{lemma}

\begin{theorem} \label{th:optimal_algorithm}
Algorithm \ref{alg:planning} will find the optimal sequence of triangles $\mathcal{S}^*$, and hence the globally optimal trajectory.

\begin{proof}
Given that Algorithm \ref{alg:planning} terminates with the optimal sequence $\mathcal{S}^*$. If we assume there exists a better sequence $\tilde{\mathcal{S}}^*$. such that:
\begin{equation*}
     Q(\boldsymbol{x}_0, \tilde{\mathcal{S}}^*, \boldsymbol{x}_f) < Q(\boldsymbol{x}_0, \mathcal{S}^*, \boldsymbol{x}_f)
\end{equation*}
Then from Lemma \ref{le:open_list_sub-sequences}, a sub-sequence $\tilde{\mathcal{S}}'$ of $\tilde{\mathcal{S}}^*$ must exist in the list of possible sequences to be extended ($open\_list$). Given the result in Lemma \ref{le:cost_lower_bound} we get that:
\begin{equation*}
    \underline{Q}(\boldsymbol{x}_0, \tilde{\mathcal{S}}', \boldsymbol{x}_f) \leq Q(\boldsymbol{x}_0, \tilde{\mathcal{S}}^*, \boldsymbol{x}_f) < Q(\boldsymbol{x}_0, \mathcal{S}^*, \boldsymbol{x}_f).
\end{equation*}
This contradicts the termination condition in \eqref{eq:termination_condition}, and hence no sequence  $\tilde{\mathcal{S}}^*$ that is better then  $\mathcal{S}^*$ can exist. 
\end{proof}
\end{theorem}

\begin{algorithm}
\caption{Optimal trajectory planning}
\label{alg:planning}
\begin{algorithmic}
    \REQUIRE Adjacency graph of triangulation, initial state $\boldsymbol{x}_0$, and goal state $\boldsymbol{x}_f$.
    \STATE $\mathcal{S}^* = [ \: ]$
    \STATE $\mathcal{S} = [\mathcal{T}_0]$ where $\boldsymbol{x}_0 \in \mathcal{T}_0$
    \STATE $open\_list = \{ \mathcal{S} \}$
    \WHILE{$open\_list$ is not empty}
    \STATE $\mathcal{S} = $ pop sequence from $open\_list$ with smallest $\underline{Q}(\boldsymbol{x}_0, \mathcal{S}, \boldsymbol{x}_f)$
    \IF{$\mathcal{S}^*$ is not empty, and $\underline{Q}(\boldsymbol{x}_0, \mathcal{S}, \boldsymbol{x}_f) \geq Q(\boldsymbol{x}_0, \mathcal{S}^*, \boldsymbol{x}_f)$}
    \RETURN Optimal triangle sequence $\mathcal{S}^*$
    \ENDIF
    \FOR{Triangle $\mathcal{T}_n$ in $neighbours(\mathcal{S})$}
    \STATE $\mathcal{S}_n = \text{extend}(\mathcal{S}, \mathcal{T}_n)$
    \IF{$\boldsymbol{x}_f \in \mathcal{T}_n$}
    \IF{$Q(\boldsymbol{x}_0, \mathcal{S}_n, \boldsymbol{x}_f) < Q(\boldsymbol{x}_0, \mathcal{S}^*, \boldsymbol{x}_f)$}
    \STATE $\mathcal{S}^* = \mathcal{S}_n$
    \ENDIF
    \ELSE
    \STATE append $\mathcal{S}_n$ to $open\_list$
    \ENDIF
    \ENDFOR
    \ENDWHILE
\end{algorithmic}
\end{algorithm}

\subsection{Implementation considerations}

\begin{figure}
    \centering
    \input{article/figures/triangle_halfplane.tikz}
    \caption{Triangle $\mathcal{T}_i$, with vertices $\boldsymbol{v}_{i, 1}$, $\boldsymbol{v}_{i, 2}$, $\boldsymbol{v}_{i, 3}$ }
    \label{fig:trinagle_halfplane}
\end{figure}
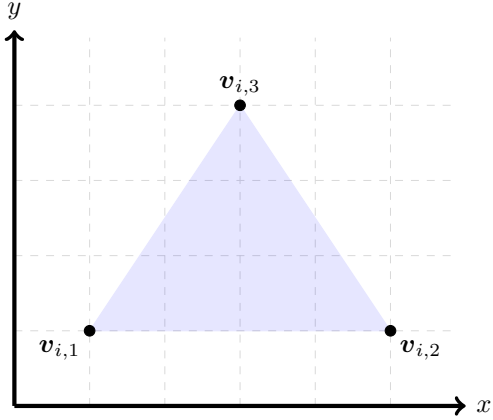

In order to implement the optimization problem given in (\ref{eq:optimization:path_refinement}), we need to formulate the constraint in (\ref{eq:optimization:path_refinement}c) as a linear inequality constraint. The most straightforward way of doing this is to use the half-space representation of the triangle. Given a 2D triangle $\mathcal{T}_i$ with vertices $\boldsymbol{v}_{i, 1}, \boldsymbol{v}_{i, 2}, \boldsymbol{v}_{i,3}$, as illustrated in Figure \ref{fig:trinagle_halfplane}, the half-space representation of a triangle gives a set of linear inequality constraints on the form:
\begin{equation*}
    \boldsymbol{A}_i \boldsymbol{p} \leq \boldsymbol{b}_i.
\end{equation*}
Where $\boldsymbol{A}_i \in \mathbb{R}^{3 \times 2}$ and $\boldsymbol{b}_i \in \mathbb{R}^{3 \times 1}$ is the matrix and vector making up the halfspace, and $\boldsymbol{p} = [x, y]^\top$ is a position. Using this, we can check if a position $\boldsymbol{p}$ lies within the triangle $\mathcal{T}_i$, as follows: 
\begin{equation}
    \boldsymbol{A}_i \boldsymbol{p} \leq \boldsymbol{b}_i \quad \Leftrightarrow \quad \boldsymbol{p} \in \mathcal{T}_i.\\
\end{equation}
The matrix $\boldsymbol{A}_i$, and vector $\boldsymbol{b}_i$ can be computed using the triangle vertices $\boldsymbol{v}_{i, 1}, \boldsymbol{v}_{i, 2}, \boldsymbol{v}_{i,3}$ as follows:
\begin{equation}
\begin{aligned}
    \boldsymbol{A}_i &=  
    \begin{bmatrix}
        (\boldsymbol{v}_{i,2} - \boldsymbol{v}_{i,1})^\top \boldsymbol{R}^\top\\
        (\boldsymbol{v}_{i,3} - \boldsymbol{v}_{i,2})^\top \boldsymbol{R}^\top\\
        (\boldsymbol{v}_{i,1} - \boldsymbol{v}_{i,3})^\top \boldsymbol{R}^\top
    \end{bmatrix} \\
    \boldsymbol{b}_i &=     
    \begin{bmatrix}
        (\boldsymbol{v}_{i,2} - \boldsymbol{v}_{i,1})^\top \boldsymbol{R}^\top \boldsymbol{v}_{i,1}\\
        (\boldsymbol{v}_{i,3} - \boldsymbol{v}_{i,2})^\top \boldsymbol{R}^\top \boldsymbol{v}_{i,2}\\
        (\boldsymbol{v}_{i,1} - \boldsymbol{v}_{i,3})^\top \boldsymbol{R}^\top \boldsymbol{v}_{i,3}
    \end{bmatrix}
\end{aligned}
\end{equation}
Where the matrix $\boldsymbol{R}$ is given as the $\pm 90^\circ$ rotation matrix, when the triangle vertices are given in a in a clockwise/counter clockwise direction. In the example in Figure \ref{fig:trinagle_halfplane}, the vertices are given in a counter clockwise direction, giving the following rotation matrix:
\begin{equation*}
    \boldsymbol{R} = 
    \begin{bmatrix}
         0 &  1 \\
        -1 &  0 
    \end{bmatrix}.
\end{equation*}

While the above linear inequality can be used to ensure the different path segments stay within the desired triangle, we propose a slight modification to this approach. The modification involves using a local triangle-centered coordinate system instead of a global coordinate system for optimizing the position within the triangle. Defining the following objects: 
\begin{equation}
\begin{aligned}
    \boldsymbol{C}_i &= 
    [\boldsymbol{v}_{i,2} - \boldsymbol{v}_{i,1}, \boldsymbol{v}_{i,3} - \boldsymbol{v}_{i,2}]\\
    \boldsymbol{d}_i &=
    \boldsymbol{v}_{i, 1},
\end{aligned}
\end{equation}
we define the transformation between the position $\boldsymbol{p} = [x, y]^\top$ in the global coordinate system, and the position $\boldsymbol{p}^\prime = [p_1^\prime, p_2^\prime]^\top$ in the local triangle coordinate system as follows:
\begin{equation}
    \boldsymbol{p} = \boldsymbol{C}_i\boldsymbol{p}^\prime + \boldsymbol{d}.
\end{equation}
Using this transformation, the triangle constraints in equation (\ref{eq:optimization:path_refinement}c) is given by the following inequality constraints:
\begin{equation}
\begin{aligned}
    0 \leq \boldsymbol{p}' &\leq 1 \\
    p_1^\prime - p_2^\prime &\leq 0.
\end{aligned}
\end{equation}
The reason for using this coordinate transformation is to help normalize the variables in the optimization problem as well as simplify the triangle constraints. This helps improve the conditioning of the optimization problem, and gives better performance when solving the problem.

Another consideration when solving (\ref{eq:optimization:path_refinement}), is how to perform the integration of the cost function (\ref{eq:optimization:path_refinement}a), and system dynamics (\ref{eq:optimization:path_refinement}b). In order to do this we propose using a multiple shooting collocation based scheme \cite{tsang1975optimal}, for which the trajectory in each triangle is approximated by a polynomial of degree $d$. This results in an optimization problem, where the objective is to find a spline where each triangle contains a polynomial representing the trajectory through the triangle (Figure \ref{fig:algorithm_sequence}d), the trajectories are then constrained to being connected between neighbouring triangles (Figure \ref{fig:algorithm_sequence}e), while at the same time satisfy the system dynamics. It is worth noting that the trajectory within each triangle will differ in length due to the size and shape of the triangle. This means the a free time variable must be used for each triangle on order in order to ensure the trajectory is constrained within the triangle.

In the graph search phase, some additional assumptions were made, in order to prune and reduce the search space.

\begin{assumption}\label{as:triangle_once}
    The optimal path will only pass through any given triangle $\mathcal{T}$ once.
\end{assumption}

Assumption \ref{as:triangle_once}, allows us to not extend a sequence of triangles into a given triangle if it already appears in the sequence. This results in a significantly smaller search space, when searching for the optimal triangle sequence. It should be noted that Assumption \ref{as:triangle_once} is not strictly necessary, as the proposed method will in theory work without it. It does however significantly reduce the search space, and helps make the method computationally feasible.

\begin{assumption}\label{as:paths_converge}
    If two initial starting points $\boldsymbol{x}_1, \boldsymbol{x}_2 \in \mathcal{T}$ are sufficiently close: 
    \begin{equation*}
        ||\boldsymbol{x}_1 - \boldsymbol{x}_2||_2 \leq \epsilon.
    \end{equation*}
    Then the optimal sequences of triangles $\mathcal{S}^*$ to the goal will be the same for both trajectories, and the difference between values of the trajectories is bounded. 
    \begin{equation*}
        ||Q(\boldsymbol{x}_1, \mathcal{S}^*, \boldsymbol{x}_f) - Q(\boldsymbol{x}_2, \mathcal{S}^*, \boldsymbol{x}_f)|| \leq \delta
    \end{equation*}
\end{assumption}

Given two different triangle sequences $\mathcal{S}_1$ and $\mathcal{S}_2$, that both end in the same triangle $\mathcal{T}$, and the same endpoints $\boldsymbol{x}_1 = \boldsymbol{x}_2, \quad \boldsymbol{x}_1, \boldsymbol{x}_2 \in \mathcal{T}$, where:
\begin{equation*}
    V(\boldsymbol{x}_0, \mathcal{S}_1) \leq V(\boldsymbol{x}_0, \mathcal{S}_2),
\end{equation*}
we only need to continue the search from the sequence $\mathcal{S}_1$, and hence can prune the sequence $\mathcal{S}_2$. Using Assumption \ref{as:paths_converge}, we can extend the above argument to say that we can prune sequences if the states are sufficiently close. Unfortunately, computing the exact bounds would require completing the trajectory, which defeats the purpose of pruning. In stead we use the following heuristic for evaluating if two endpoints $\boldsymbol{x}_1$ and $\boldsymbol{x}_2$ are sufficiently close: 
\begin{equation*}
    (\boldsymbol{x}_1 - \boldsymbol{x}_2)^\top \boldsymbol{W} (\boldsymbol{x}_1 - \boldsymbol{x}_2) \leq \epsilon, \quad \boldsymbol{x}_1, \boldsymbol{x}_2 \in \mathcal{T}
\end{equation*}
where $\boldsymbol{W}$ is a positive definite weighting matrix, and $\epsilon$ is a sufficiently small threshold. This is a relaxation of the exact condition for pruning, where $\boldsymbol{x}_1 = \boldsymbol{x}_2, \quad \boldsymbol{x}_1, \boldsymbol{x}_2 \in \mathcal{T}$, and where the conditions are exactly the same in the limit as $\epsilon \rightarrow 0$. It should be noted that pruning potential sequences is not strictly necessary. It is however added in order to further reduce the search space, and hence improve the computational complexity. 

Given algorithm \ref{alg:planning}, we can note that it is possible to paralellize the exploration of new triangle sequences. This is possible, as the exploration of possible sequences is not dependant on other sequences, however it requires some extra considerations in the termination criteria. For our implementation, this property was exploited in order to explore multiple sequences in parallel. It should be noted that if an exact heuristic function is known, the paralellization will not give a speedup, as the optimal sequence of triangles will always be the first to be explored. If however a poor heuristic function is used, parallelization will typically give a speedup, as it allows for multiple triangle sequences to be explored simultaneously. 

%% file: article/figures/algorithm_polygons.tikz
\begin{tikzpicture}[scale = 0.6]
    \draw[fill=gray, opacity=0.5] (0, 0) -- (0,  10) -- (7, 10) -- (7, 9) -- (1, 9) -- (1, 0) -- (0, 0);
    \draw[fill=gray, opacity=0.5] (3, 0) -- (10,  0) -- (10, 10) -- (9, 10) -- (9, 1) -- (3, 1) -- (3, 0);
    \draw[fill=gray, opacity=0.5] (3, 3) -- (3,  7) -- (7, 7) -- (7, 3) -- (3, 3);
\end{tikzpicture}

%% file: article/figures/algorithm_triangulation.tikz
\begin{tikzpicture}[scale = 0.6]
    \coordinate  (P0) at ( 0,  0);
    \coordinate  (P1) at ( 0, 10);
    \coordinate  (P2) at ( 7, 10);
    \coordinate  (P3) at ( 7,  9);
    \coordinate  (P4) at ( 1,  9);
    \coordinate  (P5) at ( 1,  0);
    \coordinate  (P6) at ( 3,  0);
    \coordinate  (P7) at (10,  0);
    \coordinate  (P8) at (10, 10);
    \coordinate  (P9) at ( 9, 10);
    \coordinate (P10) at ( 9,  1);
    \coordinate (P11) at ( 3,  1);
    \coordinate (P12) at ( 3,  3);
    \coordinate (P13) at ( 3,  7);
    \coordinate (P14) at ( 7,  7);
    \coordinate (P15) at ( 7,  3);

    \draw[fill=gray, opacity=0.5] (P0) -- (P1) -- (P2) -- (P3) -- (P4) -- (P5) -- (P0);
    \draw[fill=gray, opacity=0.5] (P6) -- (P7) -- (P8) -- (P9) -- (P10) -- (P11) -- (P6);
    \draw[fill=gray, opacity=0.5] (P12) -- (P13) -- (P14) -- (P15) -- (P12);
    
    \draw[color= blue, fill=blue, opacity = 0.1] (P5) -- (P6) -- (P11) -- (P5);
    \draw[color= blue, fill=blue, opacity = 0.1] (P5) -- (P12) -- (P11) -- (P5);
    \draw[color= blue, fill=blue, opacity = 0.1] (P5) -- (P4) -- (P12) -- (P5);
    \draw[color= blue, fill=blue, opacity = 0.1] (P4) -- (P13) -- (P12) -- (P4);
    \draw[color= blue, fill=blue, opacity = 0.1] (P4) -- (P13) -- (P3) -- (P4);
    \draw[color= blue, fill=blue, opacity = 0.1] (P3) -- (P13) -- (P14) -- (P3);
    
    \draw[color= blue, fill=blue, opacity = 0.1] (P9) -- (P2) -- (P3) -- (P9);
    \draw[color= blue, fill=blue, opacity = 0.1] (P9) -- (P14) -- (P3) -- (P9);
    \draw[color= blue, fill=blue, opacity = 0.1] (P9) -- (P14) -- (P10) -- (P9);
    \draw[color= blue, fill=blue, opacity = 0.1] (P10) -- (P15) -- (P14) -- (P10);
    \draw[color= blue, fill=blue, opacity = 0.1] (P10) -- (P15) -- (P11) -- (P10);
    \draw[color= blue, fill=blue, opacity = 0.1] (P12) -- (P15) -- (P11) -- (P12);
\end{tikzpicture}

%% file: article/figures/algorithm_adjacency.tikz
\begin{tikzpicture}[scale = 0.6]
    \coordinate  (P0) at ( 0,  0);
    \coordinate  (P1) at ( 0, 10);
    \coordinate  (P2) at ( 7, 10);
    \coordinate  (P3) at ( 7,  9);
    \coordinate  (P4) at ( 1,  9);
    \coordinate  (P5) at ( 1,  0);
    \coordinate  (P6) at ( 3,  0);
    \coordinate  (P7) at (10,  0);
    \coordinate  (P8) at (10, 10);
    \coordinate  (P9) at ( 9, 10);
    \coordinate (P10) at ( 9,  1);
    \coordinate (P11) at ( 3,  1);
    \coordinate (P12) at ( 3,  3);
    \coordinate (P13) at ( 3,  7);
    \coordinate (P14) at ( 7,  7);
    \coordinate (P15) at ( 7,  3);

    \draw[fill=gray, opacity=0.0] (P0) -- (P1) -- (P2) -- (P3) -- (P4) -- (P5) -- (P0);
    \draw[fill=gray, opacity=0.0] (P6) -- (P7) -- (P8) -- (P9) -- (P10) -- (P11) -- (P6);
    \draw[fill=gray, opacity=0.0] (P12) -- (P13) -- (P14) -- (P15) -- (P12);
    
    \draw[color= blue, fill=blue, opacity = 0.1] (P5) -- (P6) -- (P11) -- (P5);
    \draw[color= blue, fill=blue, opacity = 0.1] (P5) -- (P12) -- (P11) -- (P5);
    \draw[color= blue, fill=blue, opacity = 0.1] (P5) -- (P4) -- (P12) -- (P5);
    \draw[color= blue, fill=blue, opacity = 0.1] (P4) -- (P13) -- (P12) -- (P4);
    \draw[color= blue, fill=blue, opacity = 0.1] (P4) -- (P13) -- (P3) -- (P4);
    \draw[color= blue, fill=blue, opacity = 0.1] (P3) -- (P13) -- (P14) -- (P3);
    
    \draw[color= blue, fill=blue, opacity = 0.1] (P9) -- (P2) -- (P3) -- (P9);
    \draw[color= blue, fill=blue, opacity = 0.1] (P9) -- (P14) -- (P3) -- (P9);
    \draw[color= blue, fill=blue, opacity = 0.1] (P9) -- (P14) -- (P10) -- (P9);
    \draw[color= blue, fill=blue, opacity = 0.1] (P10) -- (P15) -- (P14) -- (P10);
    \draw[color= blue, fill=blue, opacity = 0.1] (P10) -- (P15) -- (P11) -- (P10);
    \draw[color= blue, fill=blue, opacity = 0.1] (P12) -- (P15) -- (P11) -- (P12);
    
    \node[circle, draw=black, inner sep=0pt] (A) at ($1/3*($(P5)+(P6)+(P11)$)$)   {A};
    \node[circle, draw=black, inner sep=0pt] (B) at ($1/3*($(P5)+(P12)+(P11)$)$)  {B};
    \node[circle, draw=black, inner sep=0pt] (C) at ($1/3*($(P5)+(P4)+(P12)$)$)   {C};
    \node[circle, draw=black, inner sep=0pt] (D) at ($1/3*($(P4)+(P13)+(P12)$)$)  {D};
    \node[circle, draw=black, inner sep=0pt] (E) at ($1/3*($(P4)+(P13)+(P3)$)$)   {E};
    \node[circle, draw=black, inner sep=0pt] (F) at ($1/3*($(P3)+(P13)+(P14)$)$)  {F};
    \node[circle, draw=black, inner sep=0pt] (G) at ($1/3*($(P9)+(P2)+(P3)$)$)    {G};
    \node[circle, draw=black, inner sep=0pt] (H) at ($1/3*($(P9)+(P14)+(P3)$)$)   {H};
    \node[circle, draw=black, inner sep=0pt] (I) at ($1/3*($(P9)+(P14)+(P10)$)$)  {I};
    \node[circle, draw=black, inner sep=0pt] (J) at ($1/3*($(P10)+(P15)+(P14)$)$) {J};
    \node[circle, draw=black, inner sep=0pt] (K) at ($1/3*($(P10)+(P15)+(P11)$)$) {K};
    \node[circle, draw=black, inner sep=0pt] (L) at ($1/3*($(P12)+(P15)+(P11)$)$) {L};
    
    \draw[<->] (A) -- (B);
    \draw[<->] (B) -- (C);
    \draw[<->] (C) -- (D);
    \draw[<->] (D) -- (E);
    \draw[<->] (E) -- (F);
    \draw[<->] (F) -- (H);
    \draw[<->] (H) -- (G);
    \draw[<->] (H) -- (I);
    \draw[<->] (I) -- (J);
    \draw[<->] (J) -- (K);
    \draw[<->] (K) -- (L);
    \draw[<->] (L) -- (B);
\end{tikzpicture}

%% file: article/figures/algorithm_spline.tikz
\begin{tikzpicture}[scale = 0.6]
    \coordinate  (P0) at ( 0,  0);
    \coordinate  (P1) at ( 0, 10);
    \coordinate  (P2) at ( 7, 10);
    \coordinate  (P3) at ( 7,  9);
    \coordinate  (P4) at ( 1,  9);
    \coordinate  (P5) at ( 1,  0);
    \coordinate  (P6) at ( 3,  0);
    \coordinate  (P7) at (10,  0);
    \coordinate  (P8) at (10, 10);
    \coordinate  (P9) at ( 9, 10);
    \coordinate (P10) at ( 9,  1);
    \coordinate (P11) at ( 3,  1);
    \coordinate (P12) at ( 3,  3);
    \coordinate (P13) at ( 3,  7);
    \coordinate (P14) at ( 7,  7);
    \coordinate (P15) at ( 7,  3);
    

    \draw[fill=gray, opacity=0.0] (P0) -- (P1) -- (P2) -- (P3) -- (P4) -- (P5) -- (P0);
    \draw[fill=gray, opacity=0.0] (P6) -- (P7) -- (P8) -- (P9) -- (P10) -- (P11) -- (P6);
    \draw[fill=gray, opacity=0.0] (P12) -- (P13) -- (P14) -- (P15) -- (P12);
    
    \draw[color= blue, fill=blue, opacity = 0.1] (P5) -- (P6) -- (P11) -- (P5);
    \draw[color= blue, fill=blue, opacity = 0.1] (P5) -- (P12) -- (P11) -- (P5);
    \draw[color= blue, fill=blue, opacity = 0.1] (P5) -- (P4) -- (P12) -- (P5);
    \draw[color= blue, fill=blue, opacity = 0.1] (P4) -- (P13) -- (P12) -- (P4);
    \draw[color= blue, fill=blue, opacity = 0.1] (P4) -- (P13) -- (P3) -- (P4);
    \draw[color= blue, fill=blue, opacity = 0.1] (P3) -- (P13) -- (P14) -- (P3);
    
    \draw[color= blue, fill=blue, opacity = 0.1] (P9) -- (P2) -- (P3) -- (P9);
    \draw[color= blue, fill=blue, opacity = 0.1] (P9) -- (P14) -- (P3) -- (P9);
    \draw[color= blue, fill=blue, opacity = 0.1] (P9) -- (P14) -- (P10) -- (P9);
    \draw[color= blue, fill=blue, opacity = 0.1] (P10) -- (P15) -- (P14) -- (P10);
    \draw[color= blue, fill=blue, opacity = 0.1] (P10) -- (P15) -- (P11) -- (P10);
    \draw[color= blue, fill=blue, opacity = 0.1] (P12) -- (P15) -- (P11) -- (P12);
    
    \coordinate (A) at ($1/3*($(P5)+(P6)+(P11)$)$)  ;
    \coordinate (B) at ($1/3*($(P5)+(P12)+(P11)$)$) ;
    \coordinate (C) at ($1/3*($(P5)+(P4)+(P12)$)$)  ;
    \coordinate (D) at ($1/3*($(P4)+(P13)+(P12)$)$) ;
    \coordinate (E) at ($1/3*($(P4)+(P13)+(P3)$)$)  ;
    \coordinate (F) at ($1/3*($(P3)+(P13)+(P14)$)$) ;
    \coordinate (G) at ($1/3*($(P9)+(P2)+(P3)$)$)   ;
    \coordinate (H) at ($1/3*($(P9)+(P14)+(P3)$)$)  ;
    \coordinate (I) at ($1/3*($(P9)+(P14)+(P10)$)$) ;
    \coordinate (J) at ($1/3*($(P10)+(P15)+(P14)$)$);
    \coordinate (K) at ($1/3*($(P10)+(P15)+(P11)$)$);
    \coordinate (L) at ($1/3*($(P12)+(P15)+(P11)$)$);
    
    \node[circle, fill=green, inner sep=2.0pt]       (S0) at (A) {};
    \node[circle, draw=gray, thick, inner sep=1.7pt] (G0) at ($ 0.6*(P5) + 0.4*(P11)$) {};
    
    \node[circle, fill=black, inner sep=1.4pt]       (S1) at ($ 0.4*(P5) + 0.6*(P11)$) {};
    \node[circle, draw=gray, thick, inner sep=1.7pt] (G1) at ($ 0.6*(P5) + 0.4*(P12)$) {};
    
    \node[circle, fill=black, inner sep=1.4pt]       (S2) at ($ 0.4*(P5) + 0.6*(P12)$) {};
    \node[circle, draw=gray, thick, inner sep=1.7pt] (G2) at ($ 0.6*(P4) + 0.4*(P12)$) {};
    
    \node[circle, fill=black, inner sep=1.4pt]       (S3) at ($ 0.4*(P4) + 0.6*(P12)$) {};
    \node[circle, draw=gray, thick, inner sep=1.7pt] (G3) at ($ 0.6*(P4) + 0.4*(P13)$) {};
    
    \node[circle, fill=black, inner sep=1.4pt]       (S4) at ($ 0.4*(P4) + 0.6*(P13)$) {};
    \node[circle, draw=gray, thick, inner sep=1.7pt] (G4) at ($ 0.6*(P3) + 0.4*(P13)$) {};
    
    
    
    \node[circle, draw=red, thick, inner sep=1.7pt]  (G7) at (G) {};
    
    \node[circle, fill=black, inner sep=1.4pt]       (S8) at ($ 0.2*(P5) + 0.8*(P11)$) {};
    \node[circle, draw=gray, thick, inner sep=1.7pt] (G8) at ($ 0.6*(P11) + 0.4*(P12)$) {};
    
    \node[circle, fill=black, inner sep=1.4pt]       (S9) at ($ 0.4*(P11) + 0.6*(P12)$) {};
    \node[circle, draw=gray, thick, inner sep=1.7pt] (G9) at ($ 0.6*(P11) + 0.4*(P15)$) {};
    
    \node[circle, fill=black, inner sep=1.4pt]       (S10) at ($ 0.4*(P11) + 0.6*(P15)$) {};
    \node[circle, draw=gray, thick, inner sep=1.7pt] (G10) at ($ 0.6*(P15) + 0.4*(P10)$) {};
    
    \path [cyan,bend left = 10 ] (S0) edge (G0) ;
    \path [cyan,bend right = 10] (S1) edge (G1);
    \path [cyan,bend left = 10 ] (S2) edge (G2);
    \path [cyan,bend right = 10] (S3) edge (G3);
    \path [cyan,bend left = 10 ] (S4) edge (G4);
    \path [cyan,bend left = 10 ] (S8) edge (G8);
    \path [cyan,bend left = 10 ] (S9) edge (G9);
    \path [cyan,bend right = 10 ] (S10) edge (G10);
    
\end{tikzpicture}

%% file: article/figures/algorithm_refinement.tikz
\begin{tikzpicture}[scale = 0.6]
    \coordinate  (P0) at ( 0,  0);
    \coordinate  (P1) at ( 0, 10);
    \coordinate  (P2) at ( 7, 10);
    \coordinate  (P3) at ( 7,  9);
    \coordinate  (P4) at ( 1,  9);
    \coordinate  (P5) at ( 1,  0);
    \coordinate  (P6) at ( 3,  0);
    \coordinate  (P7) at (10,  0);
    \coordinate  (P8) at (10, 10);
    \coordinate  (P9) at ( 9, 10);
    \coordinate (P10) at ( 9,  1);
    \coordinate (P11) at ( 3,  1);
    \coordinate (P12) at ( 3,  3);
    \coordinate (P13) at ( 3,  7);
    \coordinate (P14) at ( 7,  7);
    \coordinate (P15) at ( 7,  3);
    

    \draw[fill=gray, opacity=0.0] (P0) -- (P1) -- (P2) -- (P3) -- (P4) -- (P5) -- (P0);
    \draw[fill=gray, opacity=0.0] (P6) -- (P7) -- (P8) -- (P9) -- (P10) -- (P11) -- (P6);
    \draw[fill=gray, opacity=0.0] (P12) -- (P13) -- (P14) -- (P15) -- (P12);
    
    \draw[color= blue, fill=blue, opacity = 0.1] (P5) -- (P6) -- (P11) -- (P5);
    \draw[color= blue, fill=blue, opacity = 0.1] (P5) -- (P12) -- (P11) -- (P5);
    \draw[color= blue, fill=blue, opacity = 0.1] (P5) -- (P4) -- (P12) -- (P5);
    \draw[color= blue, fill=blue, opacity = 0.1] (P4) -- (P13) -- (P12) -- (P4);
    \draw[color= blue, fill=blue, opacity = 0.1] (P4) -- (P13) -- (P3) -- (P4);
    \draw[color= blue, fill=blue, opacity = 0.1] (P3) -- (P13) -- (P14) -- (P3);
    
    \draw[color= blue, fill=blue, opacity = 0.1] (P9) -- (P2) -- (P3) -- (P9);
    \draw[color= blue, fill=blue, opacity = 0.1] (P9) -- (P14) -- (P3) -- (P9);
    \draw[color= blue, fill=blue, opacity = 0.1] (P9) -- (P14) -- (P10) -- (P9);
    \draw[color= blue, fill=blue, opacity = 0.1] (P10) -- (P15) -- (P14) -- (P10);
    \draw[color= blue, fill=blue, opacity = 0.1] (P10) -- (P15) -- (P11) -- (P10);
    \draw[color= blue, fill=blue, opacity = 0.1] (P12) -- (P15) -- (P11) -- (P12);
    
    \coordinate (A) at ($1/3*($(P5)+(P6)+(P11)$)$)  ;
    \coordinate (B) at ($1/3*($(P5)+(P12)+(P11)$)$) ;
    \coordinate (C) at ($1/3*($(P5)+(P4)+(P12)$)$)  ;
    \coordinate (D) at ($1/3*($(P4)+(P13)+(P12)$)$) ;
    \coordinate (E) at ($1/3*($(P4)+(P13)+(P3)$)$)  ;
    \coordinate (F) at ($1/3*($(P3)+(P13)+(P14)$)$) ;
    \coordinate (G) at ($1/3*($(P9)+(P2)+(P3)$)$)   ;
    \coordinate (H) at ($1/3*($(P9)+(P14)+(P3)$)$)  ;
    \coordinate (I) at ($1/3*($(P9)+(P14)+(P10)$)$) ;
    \coordinate (J) at ($1/3*($(P10)+(P15)+(P14)$)$);
    \coordinate (K) at ($1/3*($(P10)+(P15)+(P11)$)$);
    \coordinate (L) at ($1/3*($(P12)+(P15)+(P11)$)$);
    
    \node[circle, fill=green, inner sep=2.0pt]       (S0) at (A) {};
    \node[circle, draw=gray, thick, inner sep=1.7pt] (G0) at ($ 0.32*(P5) + 0.68*(P11)$) {};
    
    \node[circle, fill=black, inner sep=1.4pt]       (S1) at (G0) {};
    \node[circle, draw=gray, thick, inner sep=1.7pt] (G1) at ($ 0.3*(P5) + 0.7*(P12)$) {};
    
    \node[circle, fill=black, inner sep=1.4pt]       (S2) at (G1) {};
    \node[circle, draw=gray, thick, inner sep=1.7pt] (G2) at ($ 0.3*(P4) + 0.7*(P12)$) {};
    
    \node[circle, fill=black, inner sep=1.4pt]       (S3) at (G2) {};
    \node[circle, draw=gray, thick, inner sep=1.7pt] (G3) at ($ 0.1*(P4) + 0.9*(P13)$) {};
    
    \node[circle, fill=black, inner sep=1.4pt]       (S4) at (G3) {};
    \node[circle, draw=gray, thick, inner sep=1.7pt] (G4) at ($ 0.5*(P3) + 0.5*(P13)$) {};
    
    \node[circle, fill=black, inner sep=1.4pt]       (S5) at (G4) {};
    
    
    \node[circle, draw=red, thick, inner sep=1.7pt]  (G7) at (G) {};
    
    \draw [cyan] plot [smooth, tension=0.5] coordinates { (S0) (S1) (S2) (S3) (S4) (S5)};
\end{tikzpicture}

%% file: article/figures/algorithm_final.tikz
\begin{tikzpicture}[scale = 0.6]
    \coordinate  (P0) at ( 0,  0);
    \coordinate  (P1) at ( 0, 10);
    \coordinate  (P2) at ( 7, 10);
    \coordinate  (P3) at ( 7,  9);
    \coordinate  (P4) at ( 1,  9);
    \coordinate  (P5) at ( 1,  0);
    \coordinate  (P6) at ( 3,  0);
    \coordinate  (P7) at (10,  0);
    \coordinate  (P8) at (10, 10);
    \coordinate  (P9) at ( 9, 10);
    \coordinate (P10) at ( 9,  1);
    \coordinate (P11) at ( 3,  1);
    \coordinate (P12) at ( 3,  3);
    \coordinate (P13) at ( 3,  7);
    \coordinate (P14) at ( 7,  7);
    \coordinate (P15) at ( 7,  3);

    \draw[fill=gray, opacity=0.0] (P0) -- (P1) -- (P2) -- (P3) -- (P4) -- (P5) -- (P0);
    \draw[fill=gray, opacity=0.0] (P6) -- (P7) -- (P8) -- (P9) -- (P10) -- (P11) -- (P6);
    \draw[fill=gray, opacity=0.0] (P12) -- (P13) -- (P14) -- (P15) -- (P12);
    
    \draw[color= blue, fill=blue, opacity = 0.1] (P5) -- (P6) -- (P11) -- (P5);
    \draw[color= blue, fill=blue, opacity = 0.1] (P5) -- (P12) -- (P11) -- (P5);
    \draw[color= blue, fill=blue, opacity = 0.1] (P5) -- (P4) -- (P12) -- (P5);
    \draw[color= blue, fill=blue, opacity = 0.1] (P4) -- (P13) -- (P12) -- (P4);
    \draw[color= blue, fill=blue, opacity = 0.1] (P4) -- (P13) -- (P3) -- (P4);
    \draw[color= blue, fill=blue, opacity = 0.1] (P3) -- (P13) -- (P14) -- (P3);
    
    \draw[color= blue, fill=blue, opacity = 0.1] (P9) -- (P2) -- (P3) -- (P9);
    \draw[color= blue, fill=blue, opacity = 0.1] (P9) -- (P14) -- (P3) -- (P9);
    \draw[color= blue, fill=blue, opacity = 0.1] (P9) -- (P14) -- (P10) -- (P9);
    \draw[color= blue, fill=blue, opacity = 0.1] (P10) -- (P15) -- (P14) -- (P10);
    \draw[color= blue, fill=blue, opacity = 0.1] (P10) -- (P15) -- (P11) -- (P10);
    \draw[color= blue, fill=blue, opacity = 0.1] (P12) -- (P15) -- (P11) -- (P12);
    
    \coordinate (A) at ($1/3*($(P5)+(P6)+(P11)$)$)  ;
    \coordinate (B) at ($1/3*($(P5)+(P12)+(P11)$)$) ;
    \coordinate (C) at ($1/3*($(P5)+(P4)+(P12)$)$)  ;
    \coordinate (D) at ($1/3*($(P4)+(P13)+(P12)$)$) ;
    \coordinate (E) at ($1/3*($(P4)+(P13)+(P3)$)$)  ;
    \coordinate (F) at ($1/3*($(P3)+(P13)+(P14)$)$) ;
    \coordinate (G) at ($1/3*($(P9)+(P2)+(P3)$)$)   ;
    \coordinate (H) at ($1/3*($(P9)+(P14)+(P3)$)$)  ;
    \coordinate (I) at ($1/3*($(P9)+(P14)+(P10)$)$) ;
    \coordinate (J) at ($1/3*($(P10)+(P15)+(P14)$)$);
    \coordinate (K) at ($1/3*($(P10)+(P15)+(P11)$)$);
    \coordinate (L) at ($1/3*($(P12)+(P15)+(P11)$)$);
    
    \node[circle, fill=green, inner sep=2.0pt]       (S0) at (A) {};
    \node[circle, draw=gray, thick, inner sep=1.7pt] (G0) at ($ 0.32*(P5) + 0.68*(P11)$) {};
    
    \node[circle, fill=black, inner sep=1.4pt]       (S1) at (G0) {};
    \node[circle, draw=gray, thick, inner sep=1.7pt] (G1) at ($ 0.3*(P5) + 0.7*(P12)$) {};
    
    \node[circle, fill=black, inner sep=1.4pt]       (S2) at (G1) {};
    \node[circle, draw=gray, thick, inner sep=1.7pt] (G2) at ($ 0.3*(P4) + 0.7*(P12)$) {};
    
    \node[circle, fill=black, inner sep=1.4pt]       (S3) at (G2) {};
    \node[circle, draw=gray, thick, inner sep=1.7pt] (G3) at ($ 0.1*(P4) + 0.9*(P13)$) {};
    
    \node[circle, fill=black, inner sep=1.4pt]       (S4) at (G3) {};
    \node[circle, draw=gray, thick, inner sep=1.7pt] (G4) at ($ 0.5*(P3) + 0.5*(P13)$) {};
    
    \node[circle, fill=black, inner sep=1.4pt]       (S5) at (G4) {};
    \node[circle, draw=gray, thick, inner sep=1.7pt] (G5) at ($ 0.9*(P3) + 0.1*(P14)$) {};
    
    \node[circle, fill=black, inner sep=1.4pt]       (S6) at (G5) {};
    \node[circle, draw=gray, thick, inner sep=1.7pt] (G6) at ($ 0.8*(P3) + 0.2*(P9)$) {};
    
    \node[circle, fill=black, inner sep=1.4pt]       (S7) at (G6) {};
    \node[circle, draw=red, thick, inner sep=1.7pt]  (G7) at (G) {};
    
    \draw [cyan] plot [smooth, tension=0.5] coordinates { (S0) (S1) (S2) (S3) (S4) (S5) (S6) (S7) (G)};
\end{tikzpicture}

%% file: article/figures/triangle_path.tikz
\begin{tikzpicture}[scale = 0.6]
    \draw[color= blue, fill=blue, opacity = 0.1] (0, 0) -- (-4,  4) -- (4, 0) -- (0, 0);
    \draw[color= blue, fill=blue, opacity = 0.1] (0, 0) -- (-4, -4) -- (4, 0) -- (0, 0);
    
    \node[circle,draw=black, fill=white, inner sep=1pt] (A) at (-1, -2) {$A$};
    \node[circle,draw=black, fill=white, inner sep=1pt] (B) at ( 1,  0) {$B$};
    \node[circle,draw=black, fill=white, inner sep=1pt] (C) at (-1,  2) {$C$};
    
    \draw[red,dashed] (A) to[out=+20,in=-90] (B);
    \draw[blue,dashed] (B) to[out=+90,in=-20] (C);
\end{tikzpicture}

%% file: article/figures/triangle_halfplane.tikz
\begin{tikzpicture}[scale = 1.0]

    \draw[help lines, color=gray!30, dashed] (0,0) grid (5.9,4.9);
    \draw[->,ultra thick] (0,0)--(6,0) node[right]{$x$};
    \draw[->,ultra thick] (0,0)--(0,5) node[above]{$y$};
    
    \coordinate  (V1) at ( 1,  1);
    \coordinate  (V2) at ( 5,  1);
    \coordinate  (V3) at ( 3,  4);
    
    \draw[color= blue, fill=blue, opacity = 0.1] (V1) -- (V2) -- (V3) -- (V1);
    
    \filldraw[] (V1) circle (2pt) node[anchor=north east] {$\boldsymbol{v}_{i, 1}$};
    \filldraw[] (V2) circle (2pt) node[anchor=north west] {$\boldsymbol{v}_{i, 2}$};
    \filldraw[] (V3) circle (2pt) node[anchor=south] {$\boldsymbol{v}_{i, 3}$};
    
\end{tikzpicture}

%% file: content/example.tex
\section{Examples} \label{sec:example}
In order to validate the method, we test it on a simple kinematic car model in in a confined environment, and compare to a Rapidly-exploring Random Tree based approach. To further prove the versatility of the method we also show it on a test scenario in trajectory planning for marine vessels in the Trondheim fjord, for which we use it to plan trajectories that minimize time, distance as well as energy. 

\subsection{Trajectory planning for a simple kinematic car model}

\subsubsection{Simple kinematic car model}
In order to verify the proposed method we will in this section show how it can be applied to planning distance optimal paths for a simple kinematic car model on the form:
\begin{equation}
    \underbrace{
    \begin{bmatrix}
        \dot{x} \\
        \dot{y} \\
        \dot{\psi}
    \end{bmatrix}
    }_{\dot{\boldsymbol{x}}}
    = 
    \underbrace{
    \begin{bmatrix}
        \cos(\psi) v \\
        \sin(\psi) v \\
        r
    \end{bmatrix}
    }_{f(\boldsymbol{x}, \boldsymbol{u})}
\end{equation}
where  $x$, $y$ is the position, $\psi$ is the heading, $v$ is the velocity, and $r$ is the turning rate. Using the constant speed $v = 1$, we are left with an under actuated system where the turning rate is the control variable $\boldsymbol{u} = r$. This type of model is often used robotics and control theory when planning paths for wheeled robots, airplanes and underwater vehicles. As the model offers a simple geometric approximation of the maneuvering capabilities of these types vehicles.

\subsubsection{Spatial constraints}
\begin{figure}
    \centering
    \begin{subfigure}{0.4\linewidth}
        \includegraphics[width=1.0\linewidth]{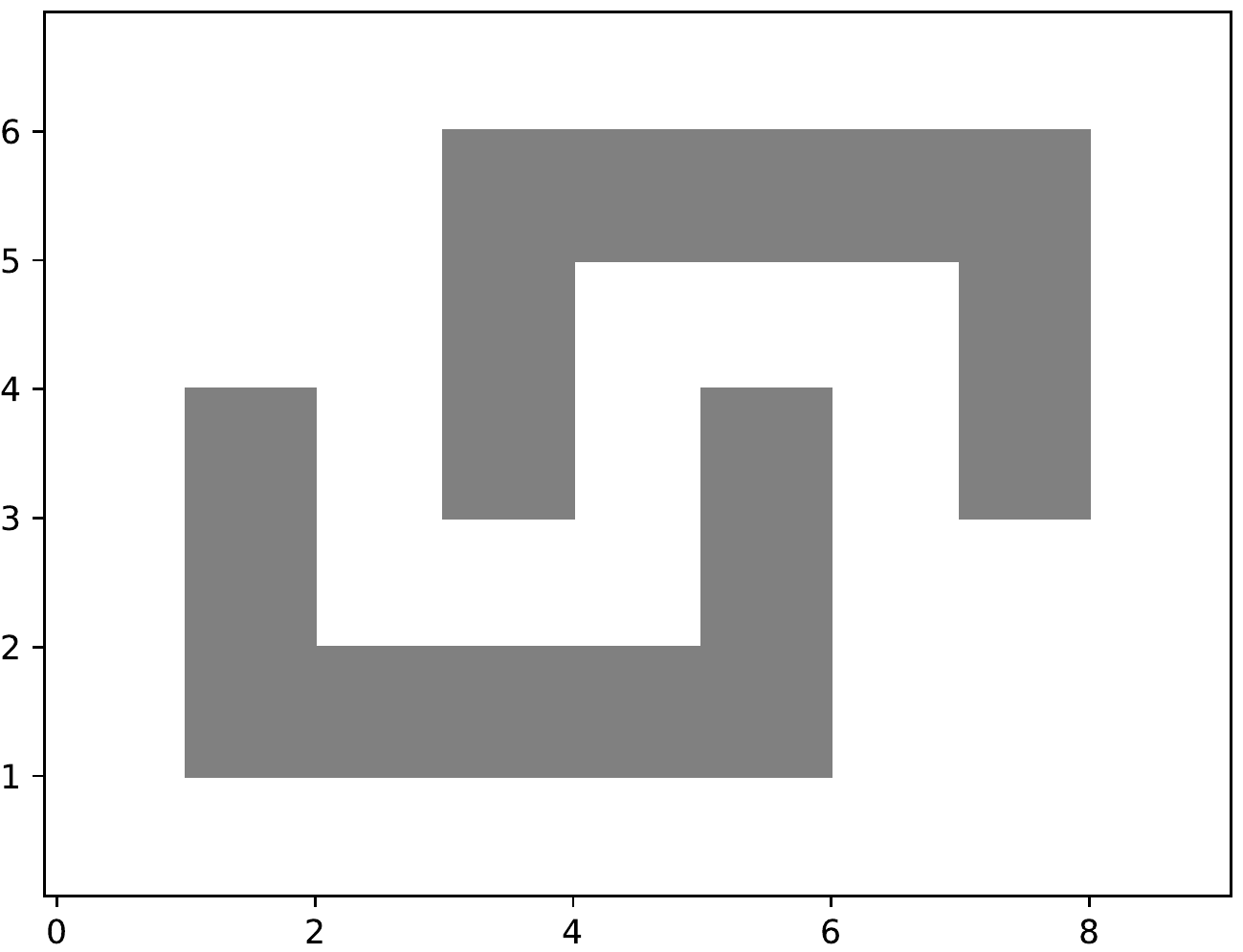}
        \caption{Constraints, without triangulation.}
    \end{subfigure}
    \begin{subfigure}{0.4\linewidth}
        \centering
        \includegraphics[width=1.0\linewidth]{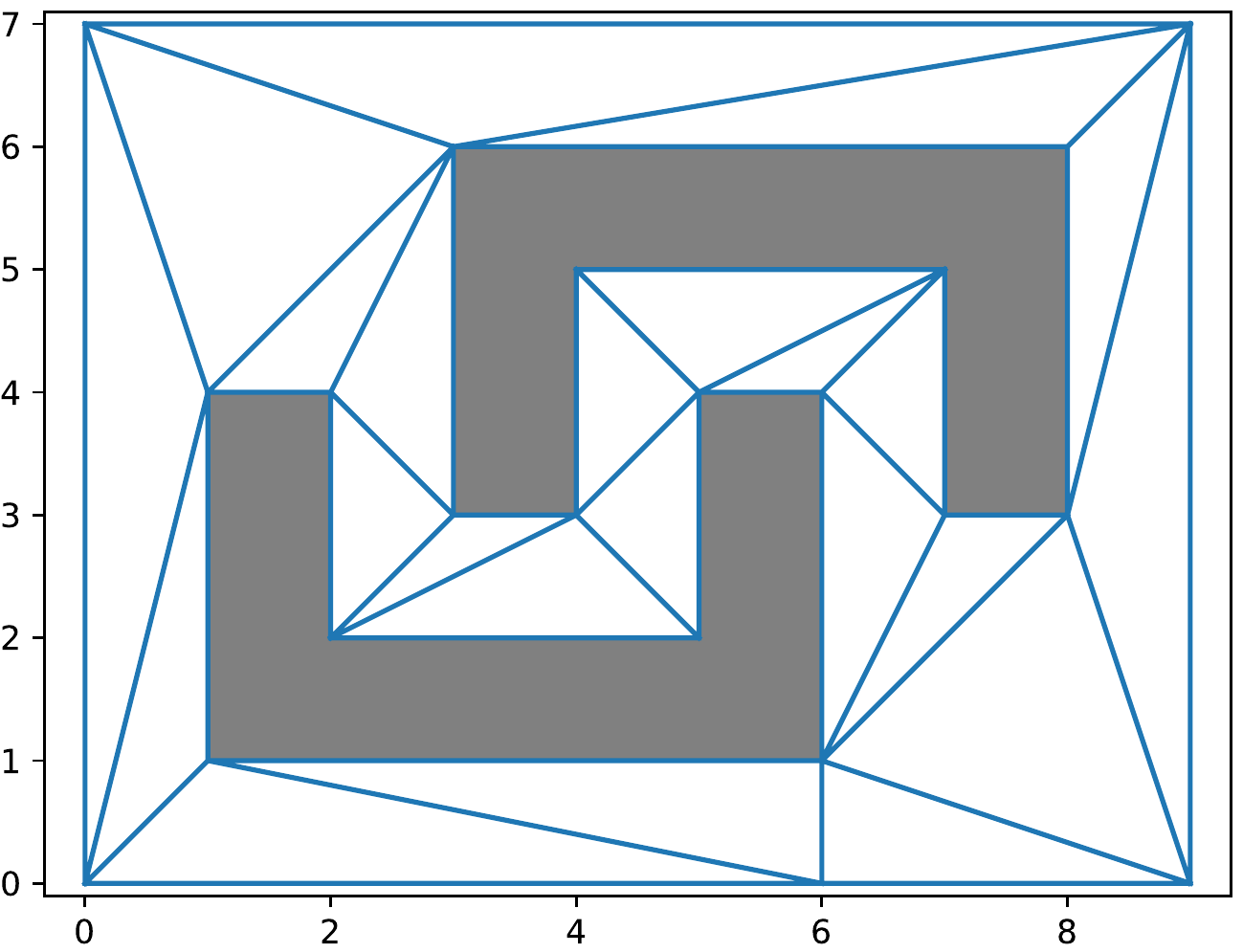}
        \caption{Constraints, with triangulation.}
    \end{subfigure}
    \caption{Polygonal spatial constraints, and the resulting CDT}
    \label{fig:conastraints_simpe_kinematic_car}
\end{figure}

In order to validate the proposed method, the simple set of spatial constraints, seen in Figure \ref{fig:conastraints_simpe_kinematic_car}a, were devised. Given the polygonal representation, the CDT was computed, giving the triangulation in Figure \ref{fig:conastraints_simpe_kinematic_car}b

\subsubsection{Optimization objective}
The objective for the optimization problem, is to find the shortest path between two points. The instantaneous is then given by the path integral as follows:
\begin{equation}
\begin{aligned}
        J(\boldsymbol{x}, \boldsymbol{u}, t) 
        &=\sqrt{ \left( \frac{dx}{dt} \right)^2 + \left( \frac{dy}{dt} \right)^2} \\
        &= \sqrt{ \left( \cos(\psi) v \right)^2 + \left( \sin(\psi) v \right)^2} \\
        &= |v| \\
        &= 1.
\end{aligned}
\end{equation}
It should be noted, that given a maximum turning rate, and the vehicle traveling at constant speed, the distance optimal path from one point to an other can be shown to be a Dubins path \cite{dubins1957curves}, which consists of straight lines and circles segments of maximum curvature.

\subsubsection{Results}
\begin{figure}
    \centering
    \begin{subfigure}{0.4\linewidth}
        \includegraphics[width=1.0\linewidth]{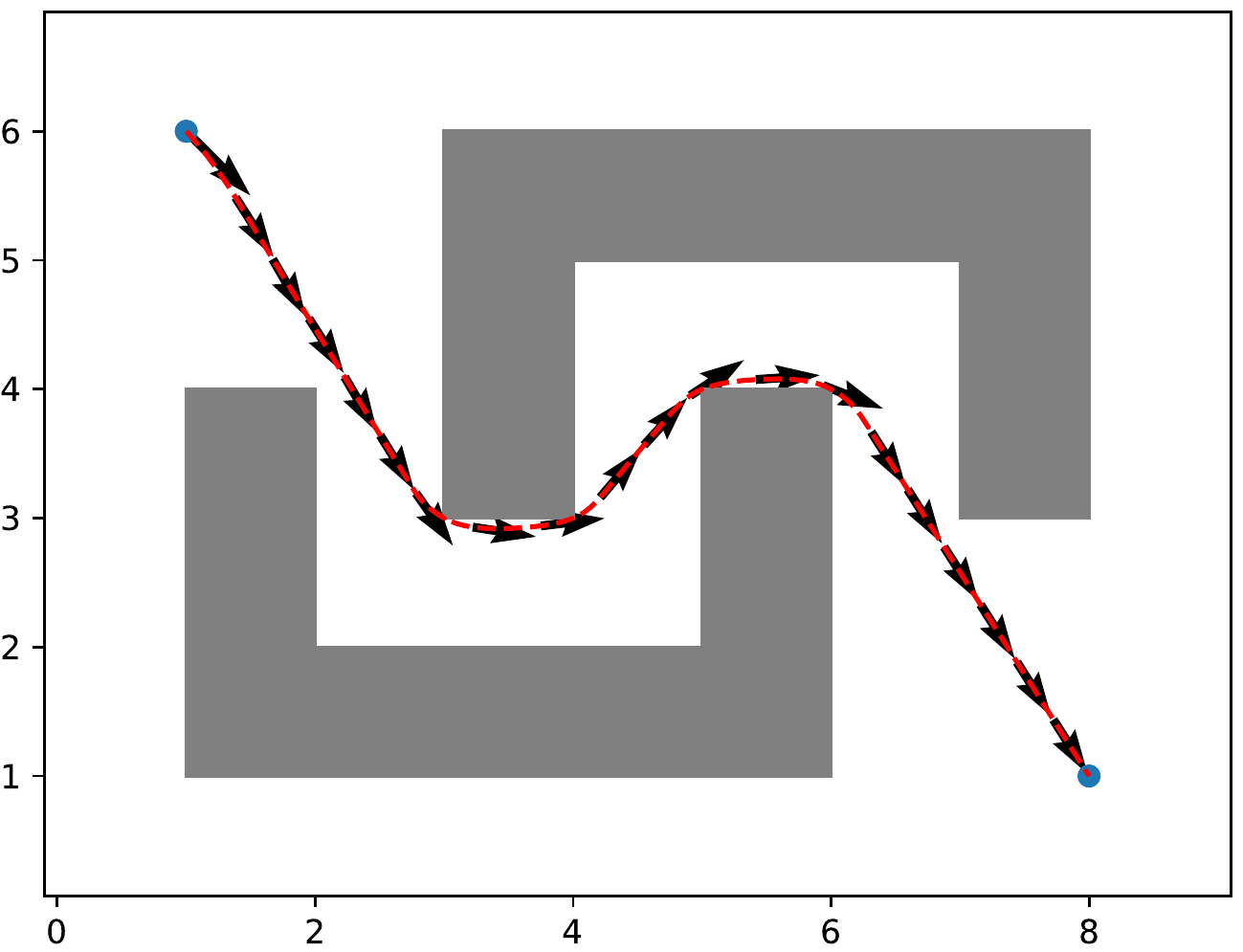}
        \caption{Our approach. The red dashed line shows the final optimized path, while the arrows show the direction of travel,  when moving between the start and end point marked with blue dots.}
    \end{subfigure}
    ~
    \begin{subfigure}{0.4\linewidth}
        \centering
        \includegraphics[width=1.0\linewidth]{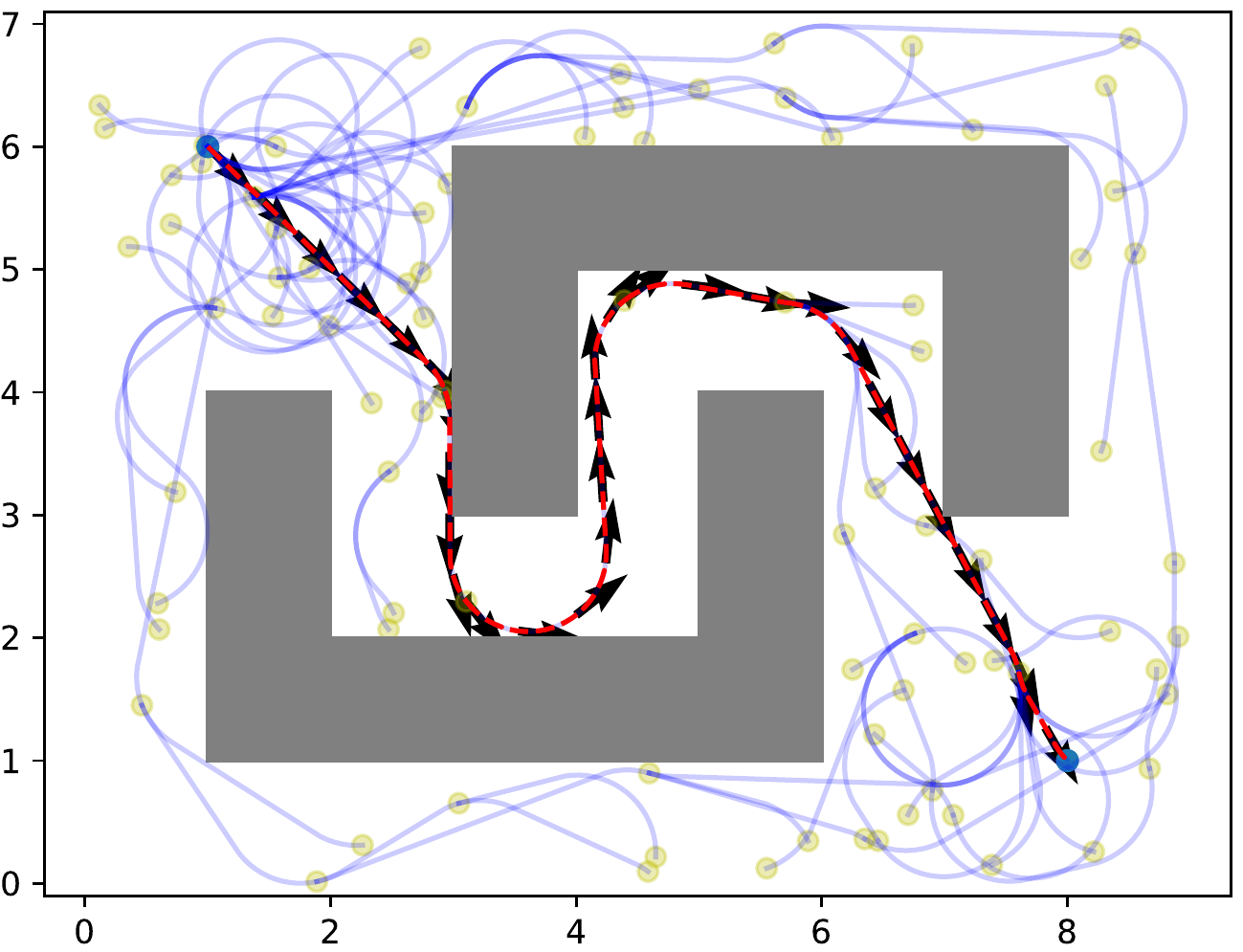}
        \caption{Dubins based RRT. The red dashed line is the final path, yellow dots show sampled nodes, while the blue lines show explored dubins paths between the sampled nodes.}
    \end{subfigure}
    \caption{Results for path generated by our proposed approach and Dubins based RRT.}
    \label{fig:results_simpe_kinematic_car}
\end{figure}

As the optimal path is known to be a Dubins path, a Dubins based RRT method \cite{vzivojevic2019path} is used for comparison, as RRT based methods are the most commonly used approaches for motion planning for robotic applications when faced with spatial constraint. Given the spatial constraint in Figure \ref{fig:conastraints_simpe_kinematic_car}, we get the resulting planned path in Figure \ref{fig:results_simpe_kinematic_car}. From the results we see that one of the major flaws of the Dubins based RRT method is that it performance is highly dependant on the randomly sampled nodes, which are used to select way points. For RRT, finding a feasible path is fairly quick, and it is possible to continue to optimize the path by generating new nodes. Further optimizing the path can often be very time consuming, as the RRT path can only guarantee converge to the optimal path as the number of sampled nodes approaches infinity \cite{karaman2011sampling}. For our proposed approach however, if a feasible path is found it is guaranteed to be optimal. This is verified in the results, where we can observe that our approach generates a path which is very similar to a Dubins path, and finds the shortest path that gets close to, but does not intersect the spatial constraints.

\subsection{Trajectory planning for an autonomous surface vessel}
In the field of marine robotics, motion planning is an important problem, which has seen a lot of interest. Given the complex vessel dynamics, as well as complex non-convex spatial constraints, the motion planning problem becomes very difficult. Because of this, most existing solutions heavily rely on simplifying the problem, this however results in loss of accuracy and optimality of the final solution. In this example we will show how our proposed planning algorithm can be used for optimal trajectory planning for a USV in the Trondheim harbour. 

\subsubsection{Vessel model}
\begin{figure}
    \centering
    \input{article/figures/3DOF_vessel.tikz}
    \caption{3-DOF vessel centered at $(x, y)$, with surge velocity $u$, sway velocity $v$, heading $\psi$ in a North-East-Down (NED) reference frame.}
    \label{fig:3DOF_vessel}
\end{figure}
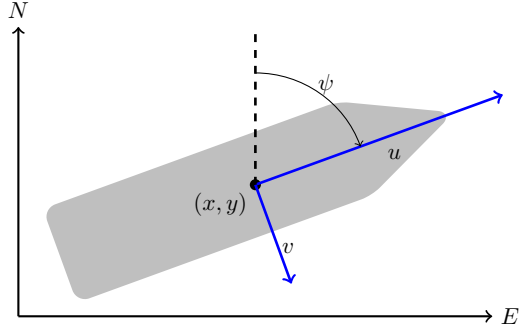

As a model for the trajectory optimization, we will use a vessel model, where we assume the vessel moves on the ocean surface in a relatively large range of possible velocities. In addition to this, we assume that the effects of the roll and pitch motions of the vessel are negligible, and hence have little impact on the surge, sway, and yaw of the vessel. The mathematical model used to describe the system can then be kept reasonably simple by limiting it to the planar position and orientation of the vessel. The motion of a surface vessel can be represented by the pose vector $\boldsymbol{\eta} = [x, y, z_r, z_i]^\top \in \mathbb{R}^4$, and the velocity vector $\boldsymbol{\nu} = [u, v, r]^\top \in \mathbb{R}^3$. Here, $(x, y)$ describe the Cartesian position in the earth-fixed reference frame, $(z_r, z_i)$ is a complex number of unit length $|z| = |z_r + i \cdot z_i| = 1$ describing the vessel orientation, where $\psi = \text{atan2}(z_i, z_r)$ is yaw angle, $(u, v)$ is the body fixed linear velocities, and $r$ is the yaw rate, an illustration is given in Figure \ref{fig:3DOF_vessel}.
Using the notation in \cite{fossen2011handbook} we can describe a 3-DOF vessel model as follows
\begin{align}
    &\dot{\boldsymbol{\eta}} = \boldsymbol{J}(\boldsymbol{\eta})\boldsymbol{\nu}, \\
    \boldsymbol{M}&\dot{\boldsymbol{\nu}} + \boldsymbol{C}(\boldsymbol{\nu})\boldsymbol{\nu} + \boldsymbol{D}(\boldsymbol{\nu})\boldsymbol{\nu} = \boldsymbol{\tau},
\end{align}
where $\boldsymbol{M}, \boldsymbol{C}(\boldsymbol{\nu}), \boldsymbol{D}(\boldsymbol{\nu}) \in \mathbb{R}^{3 \times 3}$, $\boldsymbol{\tau} \in \mathbb{R}^{3}$ and $\boldsymbol{J}(\boldsymbol{\eta})$ are the inertia matrix, coriolis matrix, dampening matrix, control input vector, and transformation matrix respectively. The transformation matrix $\boldsymbol{J}(\boldsymbol{\eta})$ is given by
\begin{equation}
   \boldsymbol{J}(\boldsymbol{\eta}) = 
   \begin{bmatrix}
    z_r & -z_i &    0 \\
    z_i &  z_r &    0 \\
      0 &    0 & -z_i \\
      0 &    0 &  z_r
   \end{bmatrix},
\end{equation}
and is the transformation from the body frame to the earth-fixed reference frame. Using the unit complex numbers in stead of a heading angle allows the the dynamics to avoid the angle wraparound problem, which avoids local optima when performing trajectory optimization. 
For the model dynamics $\boldsymbol{M}, \boldsymbol{C}(\boldsymbol{\nu}), \boldsymbol{D}(\boldsymbol{\nu})$, parameters for a simplified model of the milliAmpere experimental platform was used, where:
\begin{equation}
    \boldsymbol{M} = 
    \begin{bmatrix}
    2138 & 0 & 0 \\
    0 & 2528 & 0 \\
    0 & 0 & 3942  
    \end{bmatrix}
\end{equation}
\begin{equation}
    \boldsymbol{C}(\boldsymbol{\nu})\boldsymbol{\nu} + \boldsymbol{D}(\boldsymbol{\nu})\boldsymbol{\nu} =
    \begin{bmatrix}
    10.3 u + 114.6 |u|u - 2528vr \\
    13.0 v + 200.8 |v|v + 2138ur \\
    201.0 r + 424.1 |r|r + 390uv  
    \end{bmatrix}.
\end{equation}
For the thrust configuration, one rotatable azimuth thruster is assumed, giving the following thrust vector: 
\begin{equation}
    \boldsymbol{\tau} = 
    \begin{bmatrix}
        u_1 \cos(u_2) \\
        u_1 \sin(u_2) \\
        -2 u_1 \sin(u_2)
    \end{bmatrix},
\end{equation}
Where $0 \leq u_1 \leq 400$ is the thruster force, and $-45^\circ \leq u_2 \leq 45^\circ$ the thruster angle.

\subsubsection{Spatial constraints}
\begin{figure}
    \centering
    \begin{subfigure}{0.4\linewidth}
        \includegraphics[width=1.0\linewidth]{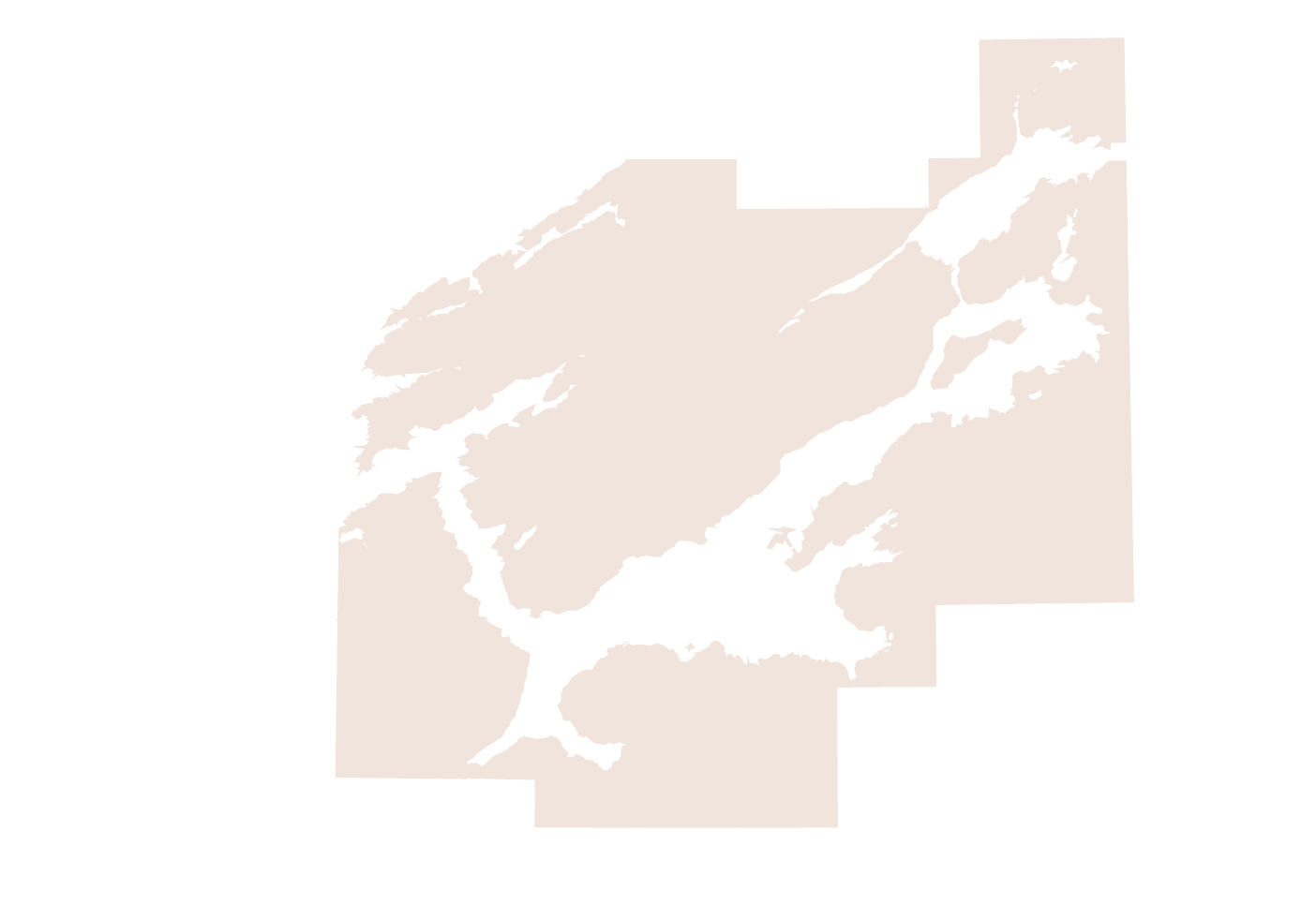}
        \caption{Map, without triangulation.}
    \end{subfigure}
    \begin{subfigure}{0.4\linewidth}
        \centering
        \includegraphics[width=1.0\linewidth]{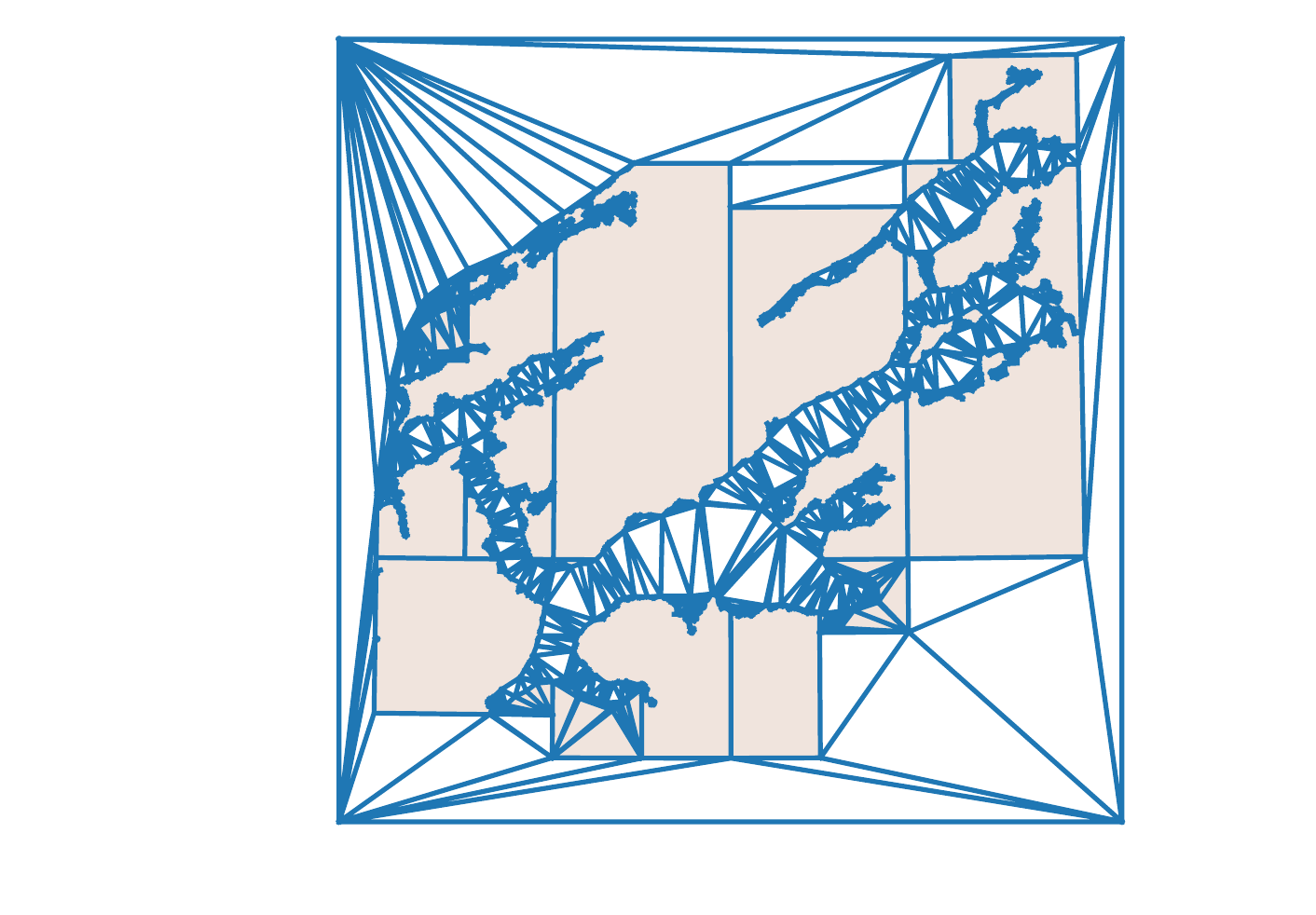}
        \caption{Map, with triangulation.}
    \end{subfigure}
    \caption{Map of the Trondheim fjord, based on polygons representing land masses.}
    \label{fig:map}
\end{figure}

Using a map where landmasses are represented by polygons, a CDT is created, where all edges of the polygons are treated as constraints. Doing this ensures that the resulting triangulation has triangles that do not intersect land. The resulting triangulation mesh is shown in Figure \ref{fig:map}. While the whole map of the Trondheim fjord is used, for the example, only a small portion of the map was relevant as the start and goal positions were selected within the Trondheim harbour.

\subsubsection{Optimization objective}
Depending on the use-case, any optimization objective satisfying Assumption \ref{as:non-negative_cost} can be selected. In this paper we will show three commonly used objectives, namely time minimization, distance minimization, and energy minimization. 

\subsubsection*{Minimum time}
In terms of instantaneous cost, the time minimization is the simplest optimization objective. where: 
\begin{equation}
    J(\boldsymbol{x}, \boldsymbol{u}, t) = 1.
\end{equation}
This gives the path integral optimization problem as follows:
\begin{equation}
    \int_{t_0}^{t_N} 1 \: dt.
\end{equation}
Minimizing the above expression then equates to minimizing the total time, $t_N - t_0$, of the the trajectory, with boundary conditions given by the initial and final state.

For the heuristic function of the minimum time, we chose the time taken traveling in a straight line from the given state $\boldsymbol{x}_N$ to the desired terminal state $\boldsymbol{x}_f$, at the maximum vessel speed $U_{\text{max}}$. Giving the following heuristic function:
\begin{equation}
    h(\boldsymbol{x}_N, \boldsymbol{x}_f) = \frac{\sqrt{(x_N - x_f)^2 + (y_N - y_f)^2}}{U_{\text{max}}}.
\end{equation}
Intuitively, we can see that this is an admissible heuristic, as it represents the time of traveling the shortest possible path, at the highest speed possible, hence it will always underestimate the time of a feasible trajectory. 

\subsubsection*{Minimum distance}
In terms of minimizing distance, we can observe that the instantaneous cost of a trajectory given by the north, and east coordinates $x(t)$ and $y(t)$ respectively, is given as the instantaneous arc length:
\begin{equation}
    \sqrt{ \left( \frac{dx}{dt} \right)^2 + \left( \frac{dy}{dt} \right)^2 }.
\end{equation}
From the kinematics we note that the square of the instantaneous cost can be rewritten as:
\begin{equation}
\begin{aligned}
    \dot{x}^2 + \dot{y}^2 
    =& \cos(\psi)^2 u^2 + \sin(\psi)^2 v^2 - \cos(\psi)\sin(\psi) u v \\
    +& \sin(\psi)^2 u^2 + \cos(\psi)^2 v^2 + \cos(\psi)\sin(\psi) u v \\
    =& (\cos(\psi)^2 + \sin(\psi)^2) (u^2 + v^2) \\
    =& u^2 +v^2,
\end{aligned}
\end{equation}
giving the following instantaneous cost.
\begin{equation}
    J(\boldsymbol{x}, \boldsymbol{u}, t) =  \sqrt{u^2 +v^2}
\end{equation}
This gives the path integral optimization problem as follows:
\begin{equation}
    \int_{t_0}^{t_N} \sqrt{u^2 +v^2} \: dt.
\end{equation}
Theoretically optimizing the above problem should give the shortest path, however for most optimization algorithms, the objective must be smooth and continuously differentiable, which is not the case when the square root is used. In order to ensure the function is continuously differentiable, a small positive number $\epsilon > 0$ is added, giving the following smooth approximation of the path integral:
\begin{equation}
    \int_{t_0}^{t_N} \sqrt{u^2 +v^2 + \epsilon} \: dt.
\end{equation}

For the heuristic function of the minimum distance, we simply chose the euclidean distance from the given state $\boldsymbol{x}_N$ to the desired terminal state $\boldsymbol{x}_f$.
\begin{equation}
    h(\boldsymbol{x}_N, \boldsymbol{x}_f) = \sqrt{(x_N - x_f)^2 + (y_N - y_f)^2}
\end{equation}
Intuitively, we can see that this is an admissible heuristic, as represents the straight line path, which is the shortest possible path between two points. This means that it will always underestimate the length of a feasible trajectory. 

\subsubsection*{Minimum energy}
In many problems, it is often useful to minimize the energy usage. In the case of marine vessels, minimizing energy usage, equates to better fuel efficiency, and less pollution. For moving objects, the quantity of work over time (power) is integrated along the trajectory of the point of application of the force. This gives the instantaneous power as the scalar product of the force/torque and the linear/angular velocity. 
\begin{equation}
    \boldsymbol{\tau}^\top \boldsymbol{\nu}
\end{equation}
In general, power regeneration and recapture is not possible for marine vessels, In order to account for this we in stead use the absolute instantaneous power, giving the following instantaneous cost:
\begin{equation}
    J(\boldsymbol{x}, \boldsymbol{u}, t) = |X \cdot u| + |Y \cdot v| + |N \cdot r|,
\end{equation}
where the thrust vector is given as $\boldsymbol{\tau} = [X, Y, N]^\top$, and velocity vector is given as $\nu = [u, v, r]$. This gives the path integral optimization problem as follows:
\begin{equation}
    \int_{t_0}^{t_N} |X \cdot u| + |Y \cdot v| + |N \cdot r| \: dt.
\end{equation}
Similarly to the minimum distance formulation, the absolute value is none smooth and the derivative not defined at $0$, in order to avoid this problem, we again use an approximation of the absolute value giving the following path integral to be optimized. 
\begin{equation}
    \int_{t_0}^{t_N} \sqrt{(X \cdot u)^2 + \epsilon} + \sqrt{(Y \cdot v)^2 + \epsilon} + \sqrt{(N \cdot r)^2 + \epsilon} \: dt.
\end{equation}

For the heuristic function of the minimum energy, it is difficult to find a good estimate for the cost to go from a given state $\boldsymbol{x}_N$ to the desired terminal state $\boldsymbol{x}_f$. Hence the heuristic:
\begin{equation}
    h(\boldsymbol{x}_N, \boldsymbol{x}_f) = 0
\end{equation}
is chosen. This is in general a poor estimate of the cost to go, and  will result in a larger number of triangles being explored, but it is an admissible heuristic and hence satisfies Assumption \ref{as:admissible_heuristic}.

\subsubsection{Results}
\begin{figure}
    \centering
    \begin{subfigure}{0.4\linewidth}
        \includegraphics[width=\linewidth]{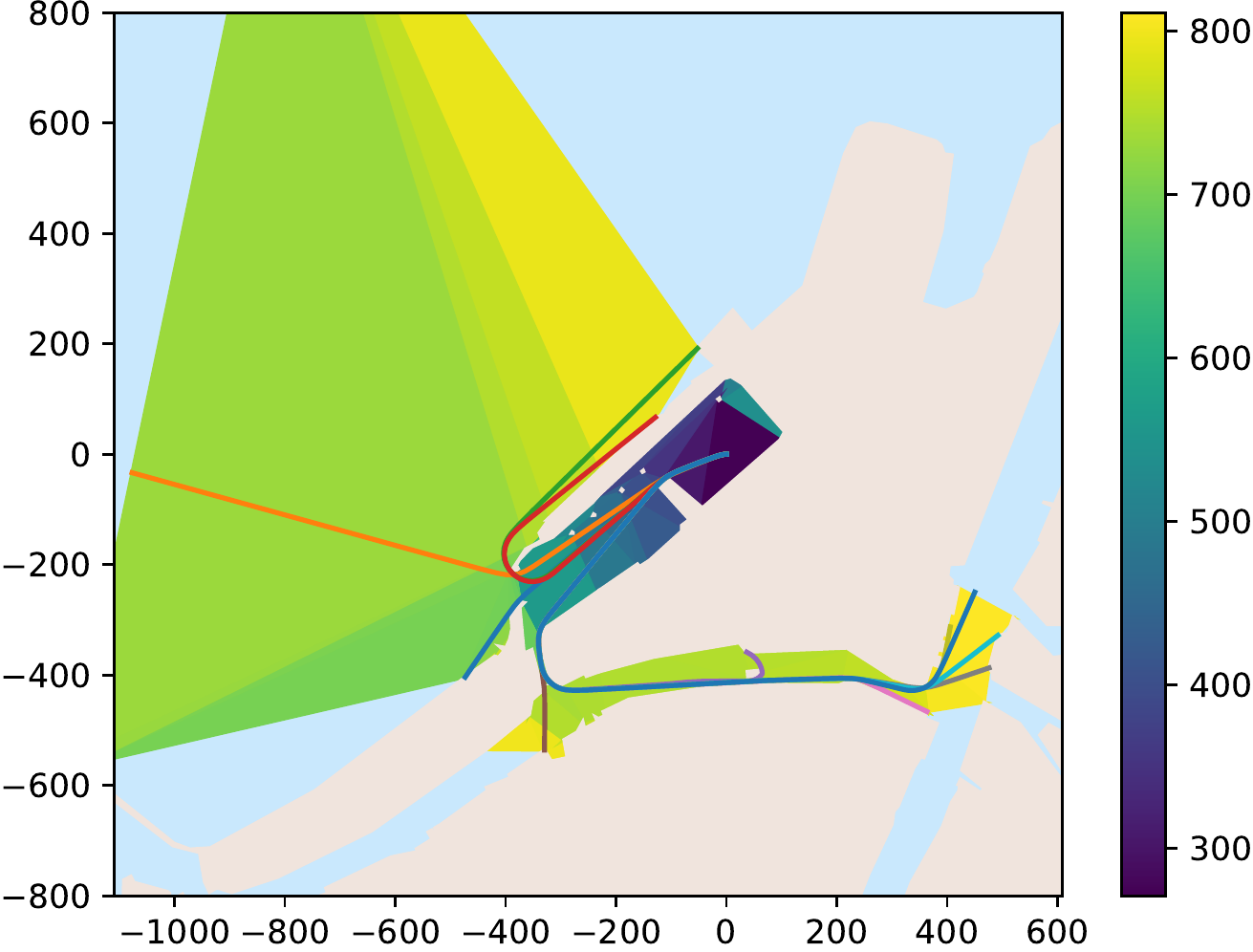}
        \caption{Space of searched triangles (colored by the value function lower bound $\underline{Q}(\cdot)$), with time optimal trajectories to the fringes given as solid lines.}
    \end{subfigure}
    ~
    \begin{subfigure}{0.4\linewidth}
        \includegraphics[width=\linewidth]{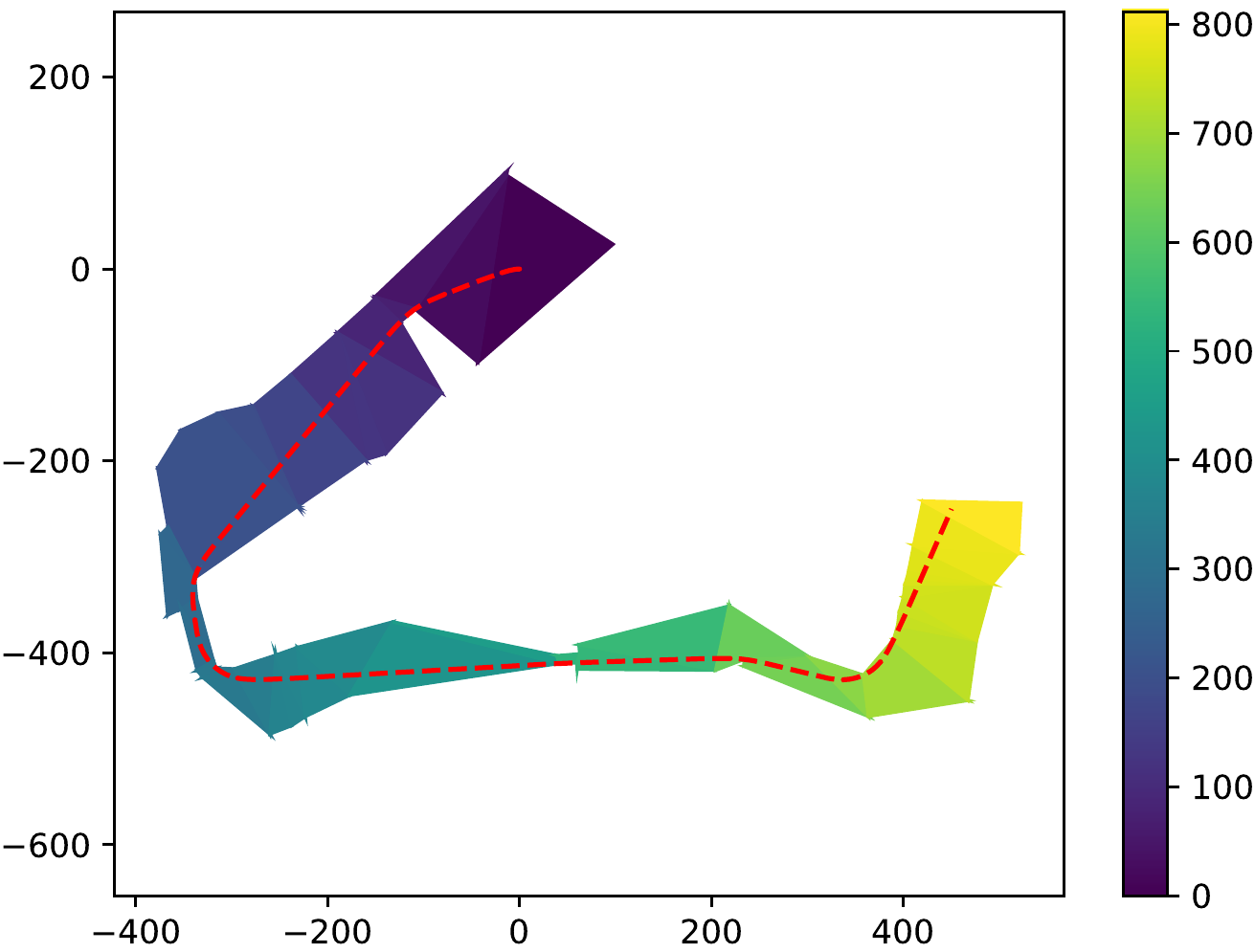}
        \caption{Sequence of triangles (colored by the free endpoint value function $V(\cdot)$) for the minimum time trajectory, within which the trajectory refinement is performed.}
    \end{subfigure}
    \caption{Search space and triangulation for minimum time trajectory.}
    \label{fig:search_illustration}
\end{figure}

\begin{figure}
    \centering
    \begin{subfigure}{0.4\linewidth}
        \includegraphics[width=\linewidth]{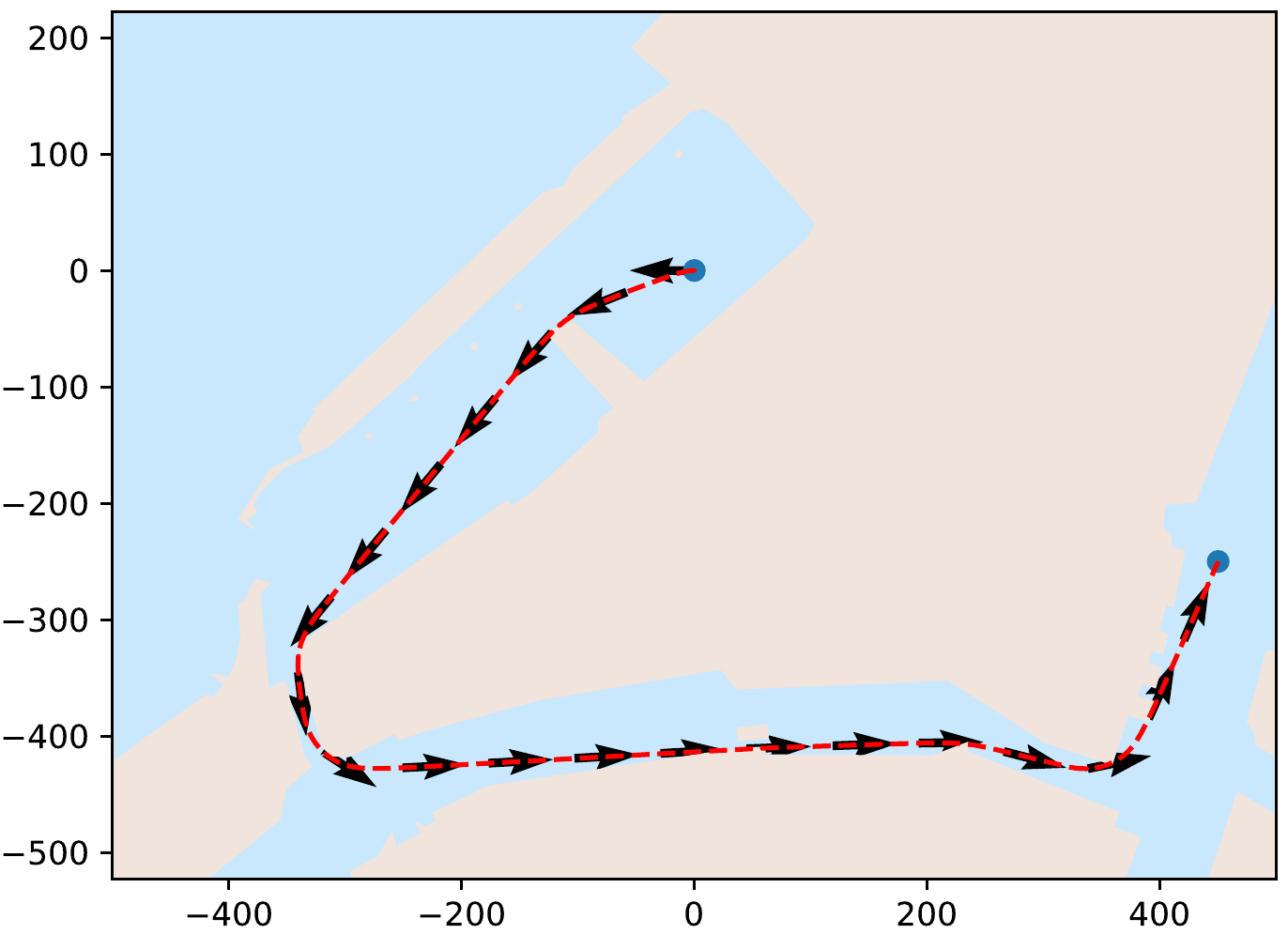}
    \end{subfigure}
    \begin{subfigure}{0.4\linewidth}
        \centering
        \resizebox{\linewidth}{!}{\input{article/figures/ma_time_optimal_state.tikz}}
    \end{subfigure}
    \caption{Minimum time path}
    \label{fig:time_optimal_path}
\end{figure}

\begin{figure}
    \centering
    \begin{subfigure}{0.4\linewidth}
        \includegraphics[width=\linewidth]{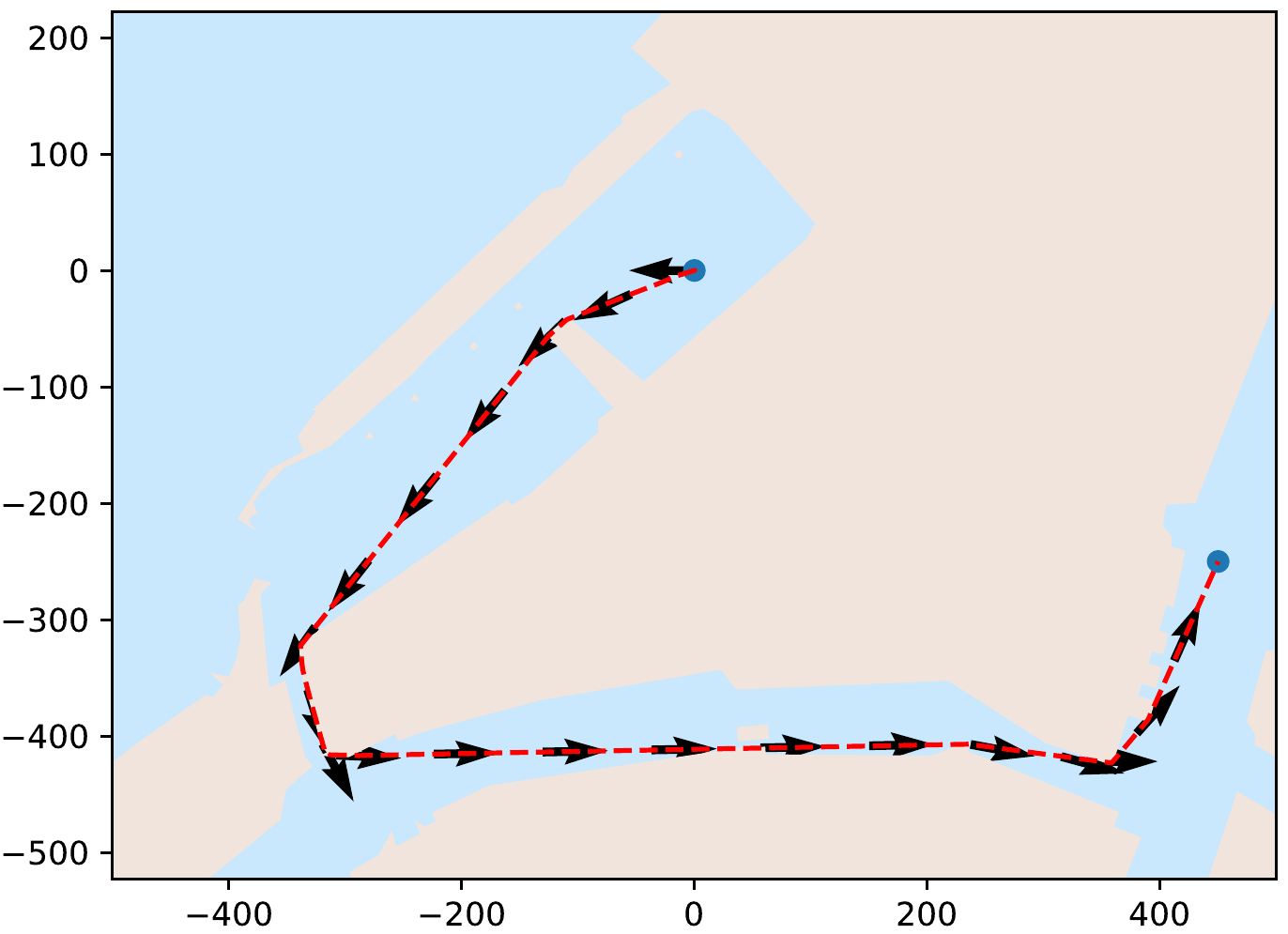}
    \end{subfigure}
    \begin{subfigure}{0.4\linewidth}
        \centering
        \resizebox{\linewidth}{!}{\input{article/figures/ma_distance_optimal_state.tikz}}
    \end{subfigure}
    \caption{Minimum distance path}
    \label{fig:distance_optimal_path}
\end{figure}

\begin{figure}
    \centering
    \begin{subfigure}{0.4\linewidth}
        \includegraphics[width=\linewidth]{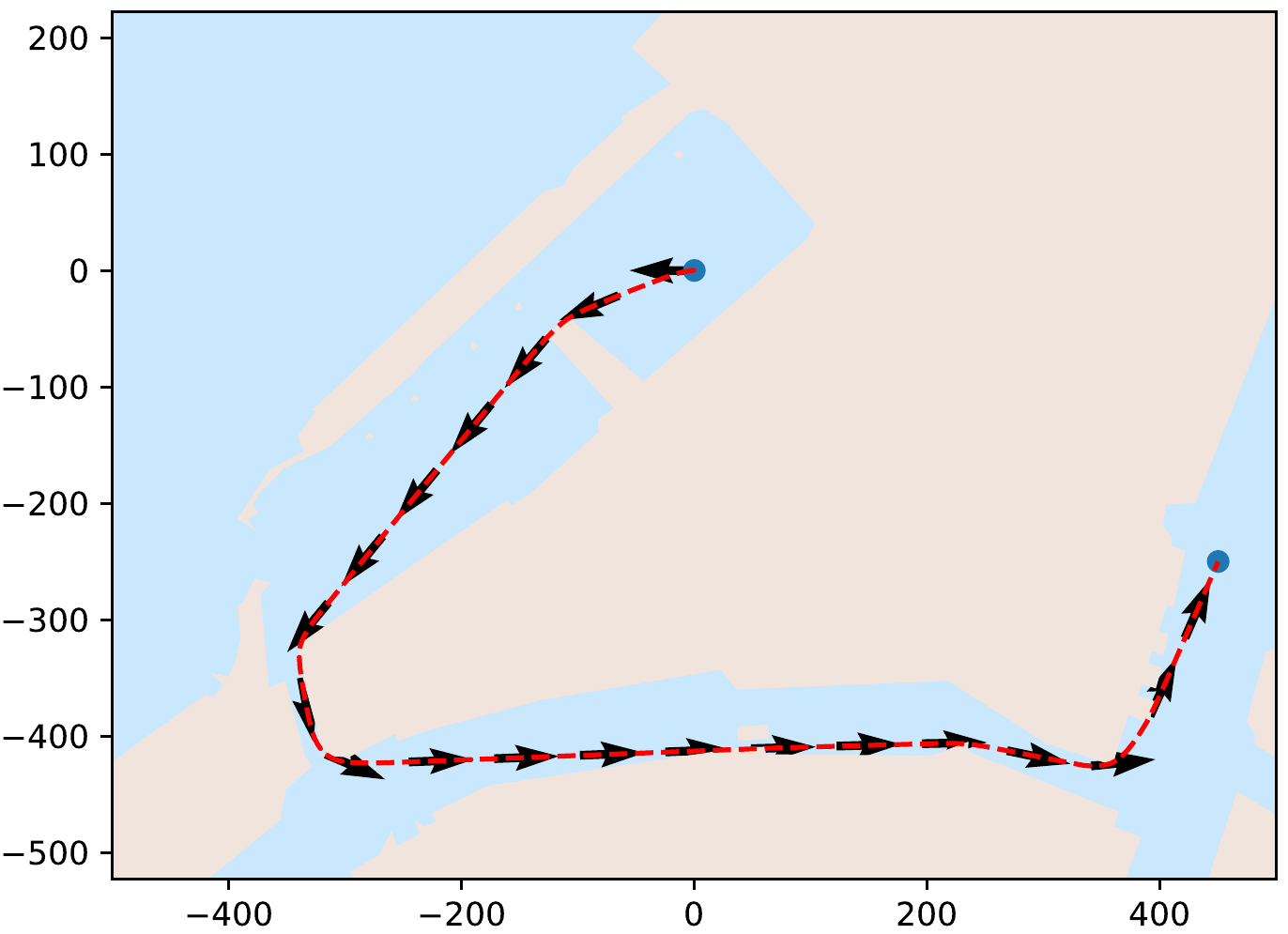}
    \end{subfigure}
    \begin{subfigure}{0.4\linewidth}
        \centering
        \resizebox{\linewidth}{!}{\input{article/figures/ma_energy_optimal_state.tikz}}
    \end{subfigure}
    \caption{Minimum energy path}
    \label{fig:energy_optimal_path}
\end{figure}

For the three different optimization objectives, the resulting trajectories from the trajectory planner are given in Figures \ref{fig:time_optimal_path}, \ref{fig:distance_optimal_path} and \ref{fig:energy_optimal_path} for the time, distance and energy minimization problems respectively. For the energy minimization problem, it is important to note that the any actuation of the control surfaces will result in energy being used, hence the optimal action would be to not move. In order to fix this, a terminal constraint was added on the time, in order to ensure the trajectory would be complete within $1200$ seconds. To visualize the proposed algorithm during the search phase, the value functions are shown in Figure \ref{fig:search_illustration}.

From the performance measure comparison in Table \ref{tab:performance_comaprison}, we can see that the different optimization objectives perform as expected, as they each minimize their respective objectives. For the minimum time objective, we can see that the speed in the surge direction is close to the maximum for most of the duration of the trajectory, this is what results in the minimum time trajectory, but comes at the cost of a slightly longer trajectory in terms of distance, and a significantly larger energy consumption. For the minimum distance trajectory we see a more erratic behaviour, especially in the surge direction. This pattern of speeding up and slowing down, is what allows the vessel to take tight corners, and hence minimize the distance, however due to the trajectory dynamics, the resulting distance is only slightly shorter than that of the other two optimization objectives. For the minimum energy trajectory, the behaviour is similar to that of the minimum time objective, with the main difference being a lower surge speed. This behaviour is due to the nonlinear drag, which makes lower speeds more energy efficient. 

\begin{figure}
    \centering
    \resizebox{0.8\linewidth}{!}{\input{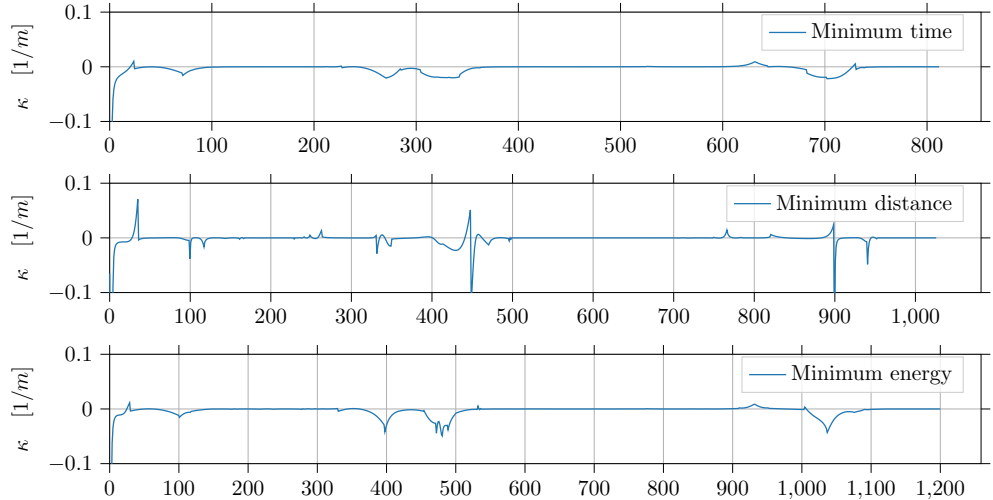}}
    \caption{Trajectory curvature resulting from the different optimization objectives.}
    \label{fig:curvature}
\end{figure}

A useful tool for evaluating a the feasibility of a trajectory, is the trajectory curvature $\kappa$.
\begin{equation}
    \kappa = \frac{\dot{x} \cdot \ddot{y} - \dot{y} \cdot \ddot{x}}{\left( \dot{x}^2 + \dot{y}^2 \right)^{\frac{3}{2}}}
\end{equation}
One of the reasons for the curvature being used as a way of evaluating trajectory feasibility, is that most vessels have a limit on the maximum possible path curvature. This has lead to the widespread use of Dubins paths \cite{dubins1957curves} which consist of straight line segments and circle arcs with maximum curvature, giving path with piecewise constant curvature. These paths have been shown to be the shortest path for a vehicle that only travels forward, and has a constraint on max curvature. The Dubins path however does not consider the underlying system dynamics, hence a dubins path is no longer optimal once the dynamics are considered. This is illustrated in Figure \ref{fig:curvature} where the curavature is continuous, similar to \cite{lekkas2013continuous}. From the curvature results it is worth noting the difference in curvature between the different optimization objectives. For the minimum time objective a higher speed is desired, hence the curvature is small allowing for taking turns at higher speeds. For the minimum distance trajectory, we can observe spikes of very high curvature, which is what we expect as the shortest path will consist only of straight line segments. For the minimum energy trajectory, we see similar results to that of the minimum time path, however the peak curvature is slightly higher, as a result of the velocities being lower.

For the implementation of Algorithm \ref{alg:planning} used to solve the trajectory planning problem, we achieved the algorithm running time given in Table \ref{tab:time_comparison}. The timing shows the results for running the algorithm sequentially, as well as the performance when running the algorithm with 4 and 8 parallel workers. In theory, increasing the number of workers, should not lead to slower running times. In practise however, there is a overhead associated with each additional worker. This is reflected in the results for the minimum time and minimum distance objectives, where the sequential approach outperforms multiple workers. For the minimum energy approach however, we see that increasing the number of workers improves the solution time. This is due to the poor choice of heuristic function, resulting in having to search a larger part of the search space, and hence the ability to evaluate multiple sequences simultaneously, outweighs the overhead of having multiple workers. It should be noted that the timing result in Table \ref{tab:time_comparison}, will vary greatly with implementation and hardware, and a more optimized implementation is likely to significantly improve the running time.

\begin{table}[]
    \centering
    \begin{tabular}{l|ccc}
         Trajectory & Time $[s]$& Distance $[m]$& Energy $[kJ]$\\ 
         \hline
         Minimum Time       & \textbf{811.81} & 1460.97 & 584.82 \\
         Minimum Distance   & 1025.82 & \textbf{1450.58} & 484.70 \\
         Minimum Energy     & 1200.00 & 1456.20 & \textbf{269.96} 
    \end{tabular}
    \caption{Performance measure}
    \label{tab:performance_comaprison}
\end{table}

\begin{table}[]
    \centering
    \begin{tabular}{l|ccc}
         & Sequential & 4 workers & 8 workers\\ 
         \hline
         Minimum Time       & \textbf{4min 28s} & 6min 1s & 7min 19s \\
         Minimum Distance   & \textbf{6min 40s} & 6min 49s & 8min 36s \\
         Minimum Energy     & 18min 36s & 15min 4s & \textbf{13min 5s} 
    \end{tabular}
    \caption{Time required for solving the different problems using the sequential approach, as well as 4 and 8 parallel workers.}
    \label{tab:time_comparison}
\end{table}

%% file: article/figures/3DOF_vessel.tikz
\tikzset{
    azimuth/.pic = { 
        \draw[gray!10, fill] (-2, 3) -- (-2, -3) -- (2, -2) -- (2, 2) -- cycle;
    }
}
\tikzset{
    vessel/.pic = {
        \draw[rounded corners, gray!50, line width=2pt, fill] (39, 0) -- (20, 9) -- (-39, 9) -- (-39, -9) -- (20, -9) -- cycle;
    }
}

\tikzset{
    tunnel/.pic = {
        \draw[gray!10, line width=2pt, fill] (-1, 4) -- (1, 3) -- (1, -3) -- (-1, -4) -- cycle;
    }
}

\begin{tikzpicture}[scale = 0.07, transform shape]

    \begin{scope}[shift={(45,25)},rotate=20]
        \coordinate (body) at (0, 0);
        
        \pic at (body) {vessel};
        
        \coordinate (u) at (50, 0);
        \coordinate (v) at (0, -20);
        \coordinate (U) at ($(u) + (v)$);

        
    \end{scope}
    
    \fill (body) circle[radius=30pt];
    \node[below left, scale = 12.5] at (body) () {$(x, y)$};
    
    \draw[->, thick] (0,0)--(90,0) node[right, scale = 12.5]{$E$};
    \draw[->, thick] (0,0)--(0,55) node[above, scale = 12.5]{$N$};
    
    \coordinate (xb) at ($ (body) + (0, 30)$);
    \draw [->,line width=1pt,color=blue] (body) -- node[below right, color=black, scale = 12.5]{$u$} (u);
    \draw [->,line width=1pt,color=blue] (body) -- node[below right, color=black, scale = 12.5]{$v$} (v);
    \draw [-,line width=1pt,color=black, dashed] (body) -- node[below, color=black]{} (xb);
    
    \pic [draw, <-, "$\psi$", angle eccentricity=1.1, angle radius=1.7cm, scale = 12.5] {angle = u--body--xb};
    
\end{tikzpicture}

%% file: content/conclusion.tex
\section{Conclusion} \label{sec:conclusion}
In this paper, we have proposed a method for planning and optimizing trajectories in an environment with static polygonal obstacles, and where the trajectories must be feasible with respect to model dynamics. Under some mild assumptions, we show that the method is able to plan globally optimal trajectories, even when faced with highly non-convex obstacles. The proposed method does however have some drawbacks. The main drawback being computational requirements, which is due to each iteration of the search phase requiring the solution of a numerical optimization problem. As well as the number of decision variables for the optimization problems increasing linearly with the number of triangles the trajectory passes through. 
Another important limitation of the proposed method is that the dynamics of the system is approximated by a single polynomial within each triangle, this can cause problems for large triangles and complex dynamical models, where the polynomial is not sufficiently rich to accurately capture the dynamics. Despite these limitations, the proposed method shows great promise based on simulation results. Offering great flexibility both in terms of environment complexity, model complexity, as well as optimization objective. 

For future work, one of the main concerns would be to improve the computational efficiency. Some potential methods for doing so, include fixing the trajectory after a certain number of triangles in order to reduce the number of decision variables at later stages, or developing better heuristics to reduce or limit the search space. Work can also be done on how to best select a numerical integration scheme to better balance accuracy, flexibility, and computational efficiency. Similarly, methods for further decomposing the triangulation may also be used to improve accuracy, especially in large triangles, or when performing complex maneuvers. It may also be interesting to add additional environmental disturbances to the problems. This would be especially useful in the case of vessel motion planning, where wind and current may greatly impact the performance.